\documentclass[aps,prr,twocolumn,titlepage,nofootinbib,linenumbers,longbibliography]{revtex4-2}
\usepackage{minitoc}
\usepackage{graphicx}
\usepackage{amsmath,amssymb}
\usepackage{hyperref,comment}
\usepackage{xcolor}
\usepackage{color}
\usepackage{ulem}
\usepackage{physics}

\usepackage{xr}
\makeatletter
\newcommand{\sitoc}{\@starttoc{sitoc}}

\makeatother

\begin{document}
\nolinenumbers
\title{\large Unifying Plasticity in Ordered and Disordered Matter\\ using Topological and Geometrical Descriptors}
\author{Xin Wang$^{1}$}
\altaffiliation{These authors contributed equally to this work.}
\author{Yang Xu$^{1,2}$}
\altaffiliation{These authors contributed equally to this work.}
\author{Jin Shang$^{1,3}$}
\author{Yi Xing$^{1}$}
\author{Jie Zhang$^{1,3}$}
\email{jiezhang2012@sjtu.edu.cn}
\author{Yujie Wang$^{1,4,5}$}
\email{yujiewang@sjtu.edu.cn}
\author{Walter Kob$^{4,6}$}
\email{walter.kob@umontpellier.fr}
\author{Matteo Baggioli$^{1,2,7}$}
\email{b.matteo@sjtu.edu.cn}
\address{$^1$School of Physics and Astronomy, Shanghai Jiao Tong University, Shanghai 200240, China}
\address{$^2$Wilczek Quantum Center, Shanghai Jiao Tong University, Shanghai 200240, China}
\address{$^3$Institute of Natural Sciences,Shanghai Jiao Tong University, Shanghai 200240, China}
\address{$^4$School of Physics, Chengdu University of Technology, Chengdu, 610059, China}
\address{$^5$State Key Laboratory of Geohazard Prevention and Geoenvironment Protection, Chengdu University of Technology, Chengdu, 610059, China}
\address{$^6$Department of Physics, University of Montpellier and CNRS, 34095 Montpellier, France}
\address{$^7$Shanghai Research Center for Quantum Sciences, Shanghai 201315, China}

\begin{abstract}
Identifying the regions responsible for plastic flow in amorphous solids remains an open problem, since structural disorder seems to prevent the direct application of concepts such as dislocations, topological defects that successfully describe irreversible deformations in crystalline systems. Here, we introduce fields of dislocation, disclination, and incompatibility densities, that in crystals reduce to the standard sources of plasticity and assess their predictive power in amorphous materials. We find that, in a simulated two-dimensional glass as well in two- and three-dimensional experimental granular systems, these fields exhibit strong spatial correlations with $D^2_{\text{min}}$, a standard descriptor used to locate plastic events  in disordered solids under shear, indicating that these fields are indeed reliable quantities for predicting plastic events. In contrast to $D^2_{\rm min}$, these fields provide deeper insight into how local rotational and translational motions become coupled well before the onset of plastic events. Moreover, by allowing the two contributions to be disentangled, they reveal that rotational defects become increasingly dominant in three dimensions. Our approach paves the way for describing plasticity in crystalline and amorphous solids in a unified manner.
\end{abstract}
\maketitle

\section{Introduction}
When subjected to external loading, solids respond through a combination of reversible elastic deformations and irreversible plastic rearrangements, which involve local structural reorganization. In crystalline materials, plasticity is understood in terms of lattice defects, primarily dislocations \cite{schmid1968plasticity,sutton2020physics}. These are topological defects characterized by a discrete Burgers vector \cite{Selinger:619845,Fumeron2023,anderson2017theory}, and can experimentally be visualized by tracking atomic positions \cite{PhysRevLett.95.225506}. Within the continuum geometric framework developed by Kondo \cite{KAZUO1964219}, Bilby \cite{doi:10.1098/rspa.1955.0171}, and Kr\"oner \cite{kroner1960general}, dislocations emerge as sources of torsion \cite{kleinert1989gauge,Kupferman2015}, with the local torsion tensor directly corresponding to the dislocation density \cite{kroner1981continuum,NYE1953153}. A closely related geometric and topological description applies to disclinations \cite{kleinert1989gauge}, defects associated with broken orientational order that play a central role in liquid crystals and also appear in crystalline solids, for instance in two-dimensional melting \cite{RevModPhys.89.040501}.

Plasticity is also ubiquitous in amorphous materials, which lack long-range order and exhibit deformation across a wide range of length scales and levels of complexity \cite{Berthier2025}. However, the absence of translational order prevents in this case a direct association of plastic activity with topological defects such as dislocations, and precludes any obvious \textit{a priori} identification of defects prior to deformation.

Building on Argon’s early work \cite{ARGON197947} on localized, irreversible atomic rearrangements, Falk and Langer \cite{PhysRevE.57.7192} introduced the observable $D^2_{\text{min}}$ to measure plastic activity in amorphous solids. This scalar quantity is based on the decoupling of local deformation into two contributions: an affine component mirroring a uniform strain field, and a non-affine component capturing local structural fluctuations. Specifically, $D^2_{\text{min}}$ measures the squared residual deviation of actual neighbor displacements from the optimal affine response. For a central reference particle $0$ at time $t$, evaluated over an interval $\Delta t$, one defines
\begin{equation} \begin{split} D^2(t,\Delta t) = \sum_n \sum_i \Bigl( & r_n^i(t) - r_0^i(t) - \sum_j (\delta_{ij} + \epsilon_{ij}) \\ & \times \left[ r_n^j(t - \Delta t) - r_0^j(t - \Delta t) \right] \Bigr)^2, \end{split} \end{equation}
where $r_n^i(t)$ and $r_0^i(t)$ are the $i$-th spatial coordinates of neighboring particle $n$ and central particle $0$, respectively, with $n$ summing over all neighbors within a chosen cutoff radius and $i,j$ indexing spatial dimensions. $\delta_{ij}$ is the Kronecker delta, while $\epsilon_{ij}$ is the local strain tensor describing the best-fit affine transformation. Minimizing $D^2(t,\Delta t)$ with respect to $\epsilon_{ij}$ isolates the non-affine contribution to the local deformation, yielding $D^2_{\min}$~\cite{PhysRevE.57.7192}. Here we use time as the evolution parameter for generality; under monotonic quasistatic deformation, the sequence of configurations can equivalently be parametrized by the accumulated strain $\gamma$, with $\Delta t$ replaced by the corresponding strain increment $\Delta\gamma$.

While the identified non-affine motion is often associated with plastic rearrangements, $D^2_{\text{min}}$ remains a phenomenological and somewhat ad hoc measure, and is not necessarily in one-to-one correspondence with irreversible plastic events. Moreover, it is largely disconnected from the geometric and topological observables used to describe defects in crystals, giving the impression that the carriers of plasticity in amorphous and crystalline solids are fundamentally distinct. However, this distinction is likely artificial, since crystals with finite disorder can be expected to smoothly interpolate between perfect crystals and glasses. Hence, it is of interest to consider quantities that allow to treat plasticity in these two extreme cases within the same framework.

In this work, we show that plastic rearrangements in amorphous solids can be formulated in terms of continuous fields that, in the crystalline limit, reduce to the standard descriptors of crystalline plasticity, namely the densities of topological defects and strain incompatibility. This approach establishes thus a unified description of plasticity across ordered and disordered solids and highlights the role of topological defects in amorphous matter \cite{baggioli2026topological}.

\section{Theory}
In this section, we define the key quantities underlying the topological density fields that characterize the plasticity of crystalline systems. We subsequently demonstrate how these concepts extend to amorphous solids.

\subsection{Crystalline solids}
In crystalline solids, the topological invariant associated with dislocations is the Burgers vector \cite{https://doi.org/10.1002/pssa.2211050139},
\begin{equation}\label{e1}
\vec{b}=\oint_{\mathcal{C}} d\vec{u},
\end{equation}
where $\mathcal{C}$ is a closed loop and $\vec{u}$ denotes the displacement field defined with respect to the ideal reference configuration,i.e. , the crystal lattice. By Stokes' theorem \cite{NYE1953153,kleinert1989gauge}, this line integral can be converted into a surface integral over any oriented surface $\mathcal{S}$ whose boundary is $\partial \mathcal{S} = \mathcal{C}$,
\begin{equation}
b_j(\vec{x}) 
= \oint_{\mathcal{C}} \partial_k u_j(\vec{x}) \, dx_k 
= \int_{\mathcal{S}} \epsilon_{ilk}\, \partial_l \partial_k u_j(\vec{x}) \, dS_i,\label{ee}
\end{equation}
where $dS_i$ is the oriented surface element normal to $\mathcal{S}$, $\epsilon_{ijk}$ is the three-dimensional Levi--Civita tensor, and the Einstein summation convention is implied. The result is independent of the specific choice of $\mathcal{S}$, provided that $\mathcal{C}$ encloses the same defect content.

The integrand in Eq.~\eqref{ee} defines the (continuous, tensorial) Nye dislocation density,
\begin{equation}\label{e2}
\alpha_{ij}(\vec{x}) = \epsilon_{ilk}\,\partial_l \partial_k u_j(\vec{x}).
\end{equation}

Similarly, the topological invariant associated with disclinations is the Frank vector \cite{deWit1973Disclinations},
\begin{equation}\label{e3}
\Omega_i=\oint_{\mathcal{C}} d\omega_i
\qquad \text{with} \qquad
\omega_i=\frac{1}{2}\epsilon_{ijk}\partial_j u_k.
\end{equation}
Applying Stokes' theorem to Eq.~\eqref{e3}, the corresponding disclination density is
\begin{align}\label{e4}
\Theta_{ij}(\vec{x})=\epsilon_{ilk}\partial_l\partial_k \omega_j(\vec{x}).
\end{align}

Importantly, within the geometric framework of mechanical deformations, the dislocation and disclination densities correspond, respectively, to the torsion and Riemann curvature tensors of the effective material metric. These definitions correctly reproduce the expected singularities associated with dislocations and disclinations in crystals (see Appendix \ref{idealsec} and \cite{kleinert1989gauge} for explicit derivation).

Finally, one can define a tensorial defect density
\begin{equation}\label{e5}
\eta_{ij}(\vec{x})=\epsilon_{ikl}\epsilon_{jmn}\partial_k \partial_m s_{ln}(\vec{x}),
\end{equation}
where the strain tensor $s_{ln}$ is related to the displacement field via
\(
s_{ln}=\tfrac{1}{2}(\partial_l u_n + \partial_n u_l).
\)
The tensor $\eta_{ij}$, i.e., the double curl of the strain, is commonly referred to as the \textit{Saint-Venant incompatibility tensor} (or \textit{incompatibility density}) and characterizes local violations of the Saint-Venant compatibility condition \cite{Love1892}. A nonzero $\eta_{ij}$ signals that the strain field cannot be derived from a single-valued displacement field, implying that the associated effective material metric cannot be globally embedded in Euclidean space. Physically, it quantifies the presence of internal (residual) stresses not induced by external loading, i.e., the stress is generated by a internal source, such as a defect. 

\begin{figure}
    \centering
    \includegraphics[width=1\linewidth]{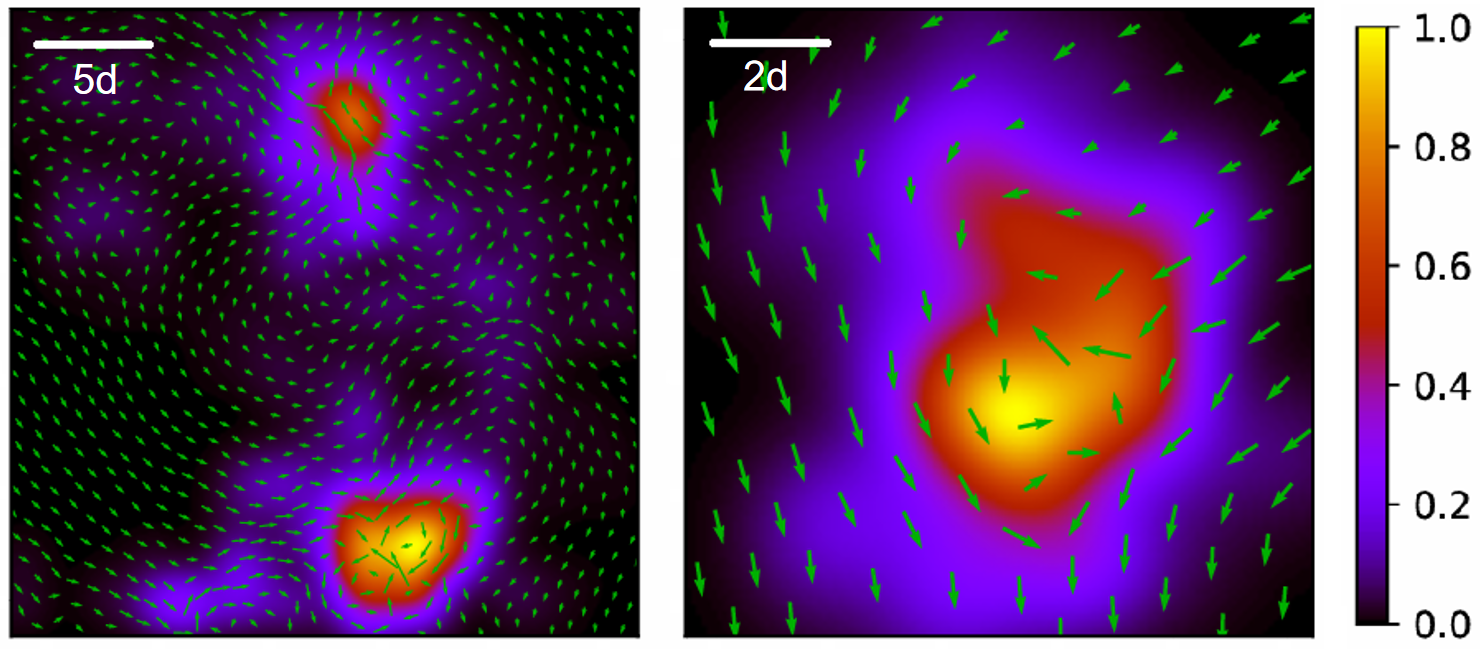}
    \caption{\textbf{Defects in the displacement field.} Representative snapshots of the particle displacement field, overlaid on a heat map of the dislocation density $|\vec{\alpha}|$, normalized by its maximum value. Color scale is given on the right. \textbf{Left panel:} 2D simulated Lennard-Jones glass with strain step $\Delta \gamma=0.5\%$. \textbf{Right panel:} Experimental 2D granular system with strain step  $\Delta \gamma=4.42\%$. In both panels, $d$ denotes the particle diameter.}
    \label{fig:1}
\end{figure}

We mention that, while dislocations and disclinations generate singularities in the incompatibility field, incompatibility is not restricted to such singular topological sources. Finite contributions can, e.g., also arise from distributed eigenstrains or non-topological defects, such as vacancies or Stone-Wales defects, which locally distort the material without introducing discrete topological charges.

Note that in two dimensions \cite{kleinert1989gauge}, the dislocation density reduces to a vector field, $\alpha_i \equiv \alpha_{3i}$, while the disclination and incompatibility densities become scalar quantities, $\Theta \equiv \Theta_{33}$ and $\eta \equiv \eta_{33}$, with the index $3$ denoting the out-of-plane direction perpendicular to the 2D system. 

\subsection{Disordered solids}
All density fields discussed in the previous section rely on defining a particle displacement field $\vec{u}(\vec{x})$. In crystalline solids, this field admits a natural definition by tracking particle positions relative to a unique, global reference frame provided by the ideal lattice structure.

However, in disordered solids, the absence of a unique, undeformed reference configuration prevents a straightforward definition of the displacement field from atomic positions. Consequently, defining a global displacement field in amorphous systems requires to make a choice on the nature of this field; moreover, unlike in crystals, it cannot be extracted from a single snapshot of particle positions. In this work, we therefore adopt the definition that $\vec{u}=\vec{x}(\gamma+\Delta\gamma)-\vec{x}(\gamma)$, evaluated over a small strain increment $\Delta\gamma$ under externally imposed shear $\gamma$, i.e., the field results from an external perturbation of the sample.

It is important to clarify the meaning of these defect-density fields in amorphous solids. Since they are constructed from a single-valued displacement field, the corresponding continuum dislocation, disclination, and incompatibility densities vanish identically when derivatives are evaluated exactly. Their finite values therefore reflect the finite-resolution reconstruction and differentiation of the discrete particle displacement field. This does not imply that the signal is physically meaningless: rather, these fields probe local departures from compatibility at the discrete coarse-grained scale and become enhanced where neighboring deformations vary sharply and cannot be smoothly matched, as it is the case around plastic rearrangements. The physically relevant information is therefore not the absolute magnitude, which vanishes upon increasing the spatial resolution, but the spatial contrast between different regions. As demonstrated in detail below and in the Appendices, this relative localization remains robust against changes in grid resolution, interpolation scheme, and grid positioning, indicating that it reflects in a genuine manner the underlying deformation structure and not just a particular numerical implementation.

It is also useful to mention a limitation of this definition relative to the crystalline case. In crystals, defects can be directly identified from the particle position, i.e., there is no need to perturb the sample. A perturbation will, however, reveal the location of the defects, since the neighboring atoms do have an enhanced mobility. Hence our use of a field generated by an external perturbation is a bona fide approach to signal the presence of defects in amorphous systems without the need of coming up with a structural definition of what exactly a defect is. However, it is important to notice that this dynamical definition cannot assess whether these defects were present before the shear stress was applied. Put another way, our method contains no information about the initial structure or its pre-existing defects.

Additionally, it is essential to emphasize that in disordered solids, quantities such as the Burgers vector, which are topologically quantized in crystals, cease to represent singular defects and instead take continuous values, corresponding to what Kl\'eman termed \textit{continuous defects} \cite{RevModPhys.80.61} (see also Gilman \cite{gilman1973flow,gilman1975mechanical}). From a formal standpoint, the continuous nature of these charges directly stems from the absence of a residual discrete spatial symmetry subgroup. Consequently, the relevant order parameter space in glasses is simply $\mathbb{R}^d$, which is topologically contractible (with trivial first homotopy group) and thus precludes quantized topological singularities. For these reasons, continuous Burgers vectors are not topological charges in the strict homotopy sense; nevertheless, as discussed below, they serve as valuable continuous fields that encode rich structural and plastic information. Consistent with this picture, continuous Burgers vectors have been shown to correlate strongly with plastic activity in glasses \cite{PhysRevE.105.024602,PhysRevLett.127.015501,liu2024measurable,bera2025burgers}. Although displacement fields in amorphous systems lack true singularities, we retain the analogy with crystals and interpret $\alpha_{ij}$, $\Theta_{ij}$, and $\eta_{ij}$ as the density fields of dislocations, disclinations, and incompatibility, respectively.

\begin{figure*}
    \centering
    \includegraphics[width=1\linewidth]{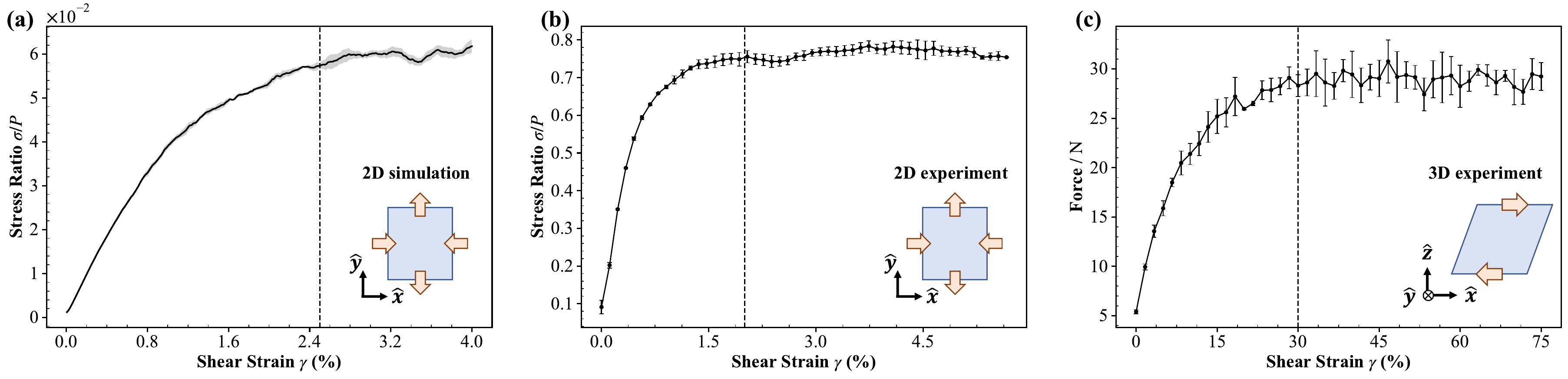}
    \caption{\textbf{Mechanical properties under shear deformation.} Stress ratio and force as a function of strain $\gamma$. \textbf{(a)} 2D simulation, \textbf{(b)} 2D experiment, and \textbf{(c)} 3D experiment. The insets show a schematic cartoon of the applied shear deformation for each case. In panels (a) and (b), $\sigma$ denotes the shear stress and $P$ the pressure. Vertical lines indicate the yielding strain. 
    }
    \label{fig:setup}
\end{figure*}

Crucially, our approach is not based on the topology of the eigenvector field, as explored in \cite{wu2023topology,Baggioli2023,Vaibhav2025,HUANG2025106274,bera2025hedgehogtopologicaldefects3d,wu2024geometrytopologicaldefectsglasses,bedollamontiel2025strikingsimilaritiesdynamicsvibrations,newNye}. Instead, it focuses on the geometrical properties of the particle displacement field, which are used to construct the density fields for disclinations, dislocations, and incompatibility. Moreover, in contrast to the vortex-like defects identified in displacement fields in \cite{10.1093/pnasnexus/pgae315,PhysRevE.109.L053002,PhysRevB.110.014107,wang2025topological,bera2025microscopic}, we employ measures that naturally recover the classical notions of dislocations and disclinations in crystalline solids. Based on this continuity perspective, this construction avoids the ambiguities associated with defining topological defects in glasses.

To illustrate the meaning of the quantities introduced above, Fig.~\ref{fig:1} presents two representative examples of the displacement field, together with the corresponding heat map of the dislocation density (see Appendix \ref{appcomp} for details about the computational methods and Appendix \ref{appB} for additional cases). As highlighted by the yellow/orange regions, the density accurately identifies incompatible zones, including local dislocation-like slip patterns (top spot in the left panel) and more complex geometrically incompatible structures (bottom spot in the left panel and the dominant signal in the right panel). By contrast, aside from minor numerical noise (blue regions), $|\vec{\alpha}|$ is small  where the displacement field remains smooth. Importantly, regions characterized by large gradients yet a smooth underlying vector field do have a small value of the dislocation density, as they do not represent defects in this sense. 

In Appendix \ref{appB}, we include a comparison with vortex-like defects identified via the winding number, following the original construction applied to the eigenvector field in \cite{wu2023topology}. This analysis reveals a pronounced correlation between the location of the vortex defects and the spatial structure of the Nye dislocation density.

\section{Systems}
To investigate how well the defect density fields detect plasticity, we consider three systems: (i) A 2D binary Lennard–Jones glass with $10^5$ particles, (ii) A 2D bidisperse frictional granular system consisting of $4065$ photoelastic disks, and (iii) A monodisperse 3D frictional granular system made of $4200$ 3D-printed
plastic spheres.  
The 2D systems (both numerical and experimental) are driven by quasistatic pure shear applied through moving walls, while the 3D system is subjected to quasistatic simple shear. 
\subsection{Simulated 2D Lennard-Jones glass} 
The system is a binary Lennard--Jones model that is not prone to crystallization \cite{ka2008glass}. The interaction between particles of species $\alpha,\beta\in\{A,B\}$ is given by
\begin{equation}
V_{\alpha\beta}(r)=4\epsilon_{\alpha\beta}
\left[
\left(\frac{d_{\alpha\beta}}{r}\right)^{12}
-
\left(\frac{d_{\alpha\beta}}{r}\right)^6
\right],
\label{eq:KA}
\end{equation}
with
\(
\epsilon_{AA}=1.0,\,
\epsilon_{AB}=1.5,\,
\epsilon_{BB}=0.5
\)
and
\(
d_{AA}=1.0,\,
d_{AB}=0.8,\,
d_{BB}=0.88
\).
We use $N=10^5$ particles with a composition of $N_A:N_B=$65:35. The system is initialized at a temperature $T=10^{-5}$ in a square box with periodic boundary conditions. Four confining walls are generated by exploiting the periodic boundary conditions and freezing a particle layer of width $3\,d_{AA}$ around the central simulation box.

\begin{figure*}
    \centering
    \includegraphics[width=\linewidth]{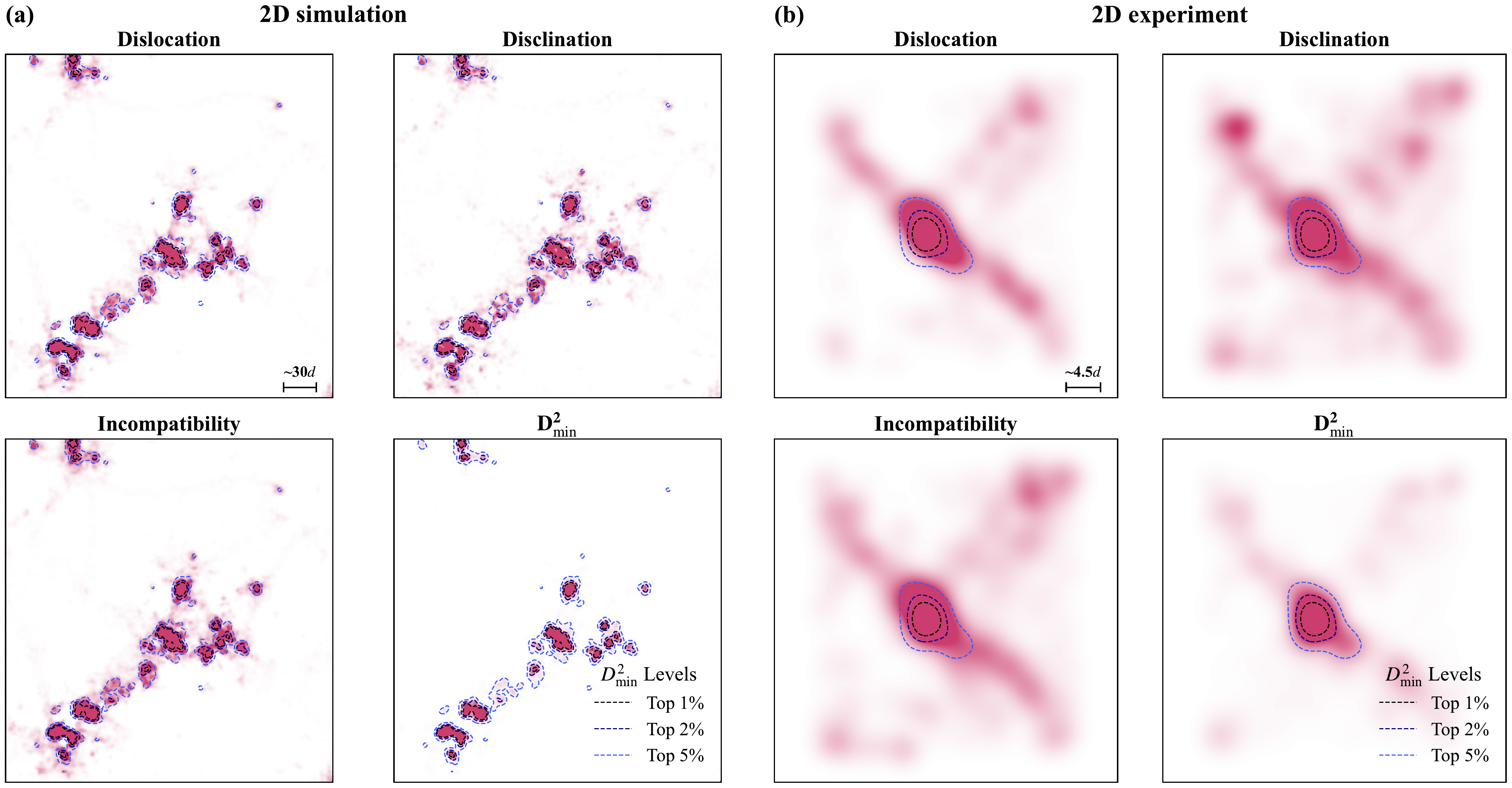}
    \caption{\textbf{Comparison between defect densities and $D^{2}_{\text{min}}$ in 2D.} Heat maps of the (normalized) dislocation, Eq.~\eqref{e2}, disclination, Eq.~\eqref{e4}, and incompatibility, Eq.~\eqref{e5} densities, together with $D^{2}_{\text{min}}$. For the dislocation density, we plot $|\vec{\alpha}|$. \textbf{(a)} 2D Lennard-Jones glass at strain $\gamma=3.38\%$ with $\Delta \gamma = 0.02\%$. \textbf{(b)} 2D experimental granular system at $\gamma=4.64\%$ with $\Delta \gamma = 0.113\%$. In all panels, the dashed contours indicate the top-percentile clusters of $D^2_{\text{min}}$. In both systems, $d$ refers to the average particle diameter.}
    \label{fig:2}
\end{figure*}

To obtain a well annealed, low-temperature glass, the system is quenched to the temperature $T=10^{-7}$ in the $NVT$ ensemble within $10^6$ steps. The box size was 280.3 which corresponds to a particle density of 1.27. Subsequently, one pair of walls is driven at constant velocity $|v_x| \approx 10^{-3}$ to impose a shear rate $\dot{\gamma} \approx 10^{-5}$, while the opposite walls move to preserve the area of the bulk region, mirroring the experimental protocol of pure shear. The resulting stress--strain response is shown in Fig.~\ref{fig:setup}(a). The inset illustrates the pure shear deformation applied. The yielding transition, estimated from the strain at which the stress starts to become flat, is estimated to be located at $\gamma \approx 2.5\%$. In the similarity analysis below, we excluded approximately six particle layers near the walls and focused on the bulk region.

\subsection{Experimental 2D granular matter system}  
The 2D granular assembly consists of $2710$ small and $1355$ large photoelastic disks with diameters $d_s = 10\ \mathrm{mm}$ and $d_l = 14\ \mathrm{mm}$. The disks rest on a horizontal glass plate inside a rectangular region bounded by two pairs of independently movable smooth walls, enabling area-preserving pure shear. Experimental data were recorded using a $2 \times 2$ array of high-resolution cameras (10 pixels/mm) mounted above the system. Additional details on the setup can be found in Refs.~\cite{wang2022experimental,Wang2021PRR}. The stress–strain curve for this system is shown in Fig.~\ref{fig:setup}(b). The inset depicts the applied pure shear deformation. The yielding transition is located at $\gamma \approx 2\%$. In the experiment, the uncertainty in particle position determination is on the order of $10^{-2}\,d_s$. In the similarity analysis below, we excluded approximately six particle layers near the walls and focused on the bulk region.

\subsection{Experimental 3D granular matter system}  
The 3D granular packing consists of monodisperse, 3D-printed plastic spheres of diameter $d = 6\ \mathrm{mm}$. The shear cell has dimensions $24d$ times $24d$ time $24d$ and is sheared in the $x$-direction. More than $4200$ particles located at least $2.5 \,d$ away from the shear cell boundaries were analyzed. The packing was first cyclically sheared under constant pressure to reach a reproducible initial volume fraction, and then sheared unidirectionally in uniform strain increments until the sample attained a maximum in volume, i.e., it has reached the state of random loose packing. X-ray tomography was used to capture the 3D structure at each strain step, while a force sensor simultaneously measured the shear force. Further experimental details are provided in Ref.~\cite{xing2024origin}. The resulting stress–strain curve is shown in Fig.~\ref{fig:setup}(c). The inset illustrates the simple shear deformation applied. The yielding transition is located at $\gamma \approx 30\%$. In this experiment, the uncertainty the determination of the particle position is on the order of $10^{-3}\,d$. In the similarity analysis below, we excluded approximately four particle layers near the walls and focused on the bulk region.

\section{Continuous defect densities and plastic activity} 

We start our analysis from the particle displacement field $u(\vec{x})$, computed at each strain step. To evaluate the densities defined in Eqs.~\eqref{e2}--\eqref{e5}, $u(\vec{x})$ is mapped onto a square (or cubic) lattice with spacing $a = d$, where $d$ denotes the average particle diameter. We have verified that the results presented in the following are robust against variations in grid size, interpolation scheme, and even externally added noise. Additional details are provided in the Appendices \ref{apps1} and \ref{apps2}.

Figure~\ref{fig:2}(a) shows, for the 2D Lennard--Jones system, the normalized density maps of the dislocations, Eq.~\eqref{e2}, disclinations, Eq.~\eqref{e4}, and incompatibilities, Eq.~\eqref{e5}, together with the $D^2_{\text{min}}$, all measured at strain $\gamma=3.38\%$ with $\Delta \gamma = 0.020\%$. Additional snapshots are provided in Appendix \ref{appmore}. A striking visual similarity emerges among all four fields, particularly within the top-percentile clusters of $D^2_{\text{min}}$ highlighted by dashed contours. This indicates that the three vector fields do carry similar information as $D^2_{\rm min}$, and hence are useful indicators for plastic activity. A more detailed analysis of this point is provided below.

Panel (b) shows the same analysis for the 2D experimental system at $\gamma = 4.64\%$ with $\Delta \gamma = 0.113\%$. A clear correspondence between the four fields is again observed, although all fields appear noisier. We have checked that this increased noise is due to the smaller length scale shown in the experimental data, i.e., a similar magnitude of noise is also observed in the simulation data if one zooms in more (see Appendix \ref{appblur}). However, especially when focusing on the top-percentile regions, the agreement between the different measures remains strong.

\begin{figure}
    \centering
    \includegraphics[width=\linewidth]{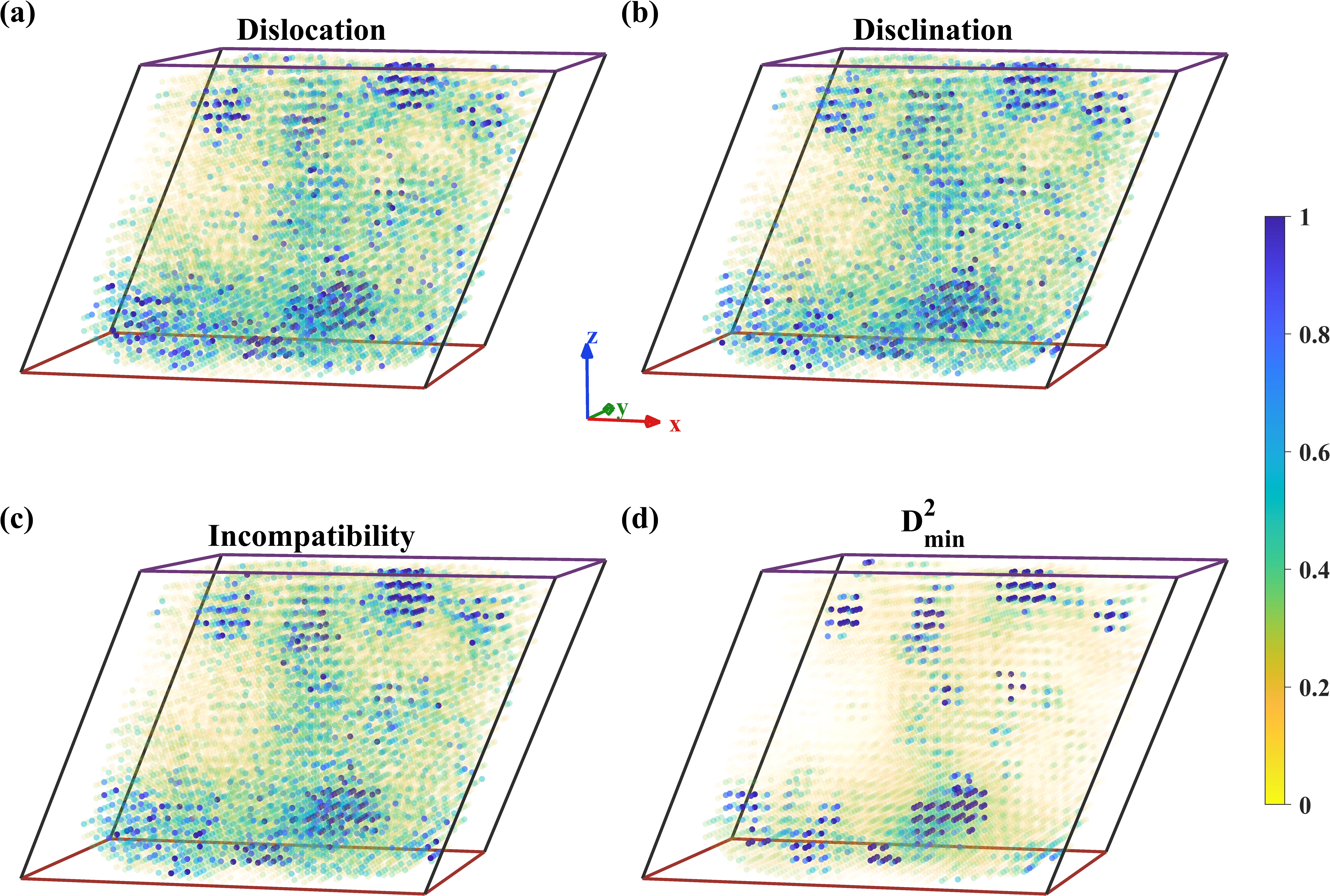}
    \caption{\textbf{Comparison between defect densities and $D^{2}_{\text{min}}$ in 3D.} Normalized norm of the \textbf{(a)} dislocation, \textbf{(b)} disclination, \textbf{(c)} incompatibility densities and \textbf{(d)} $D^2_{\text{min}}$ in 3D experimental granular system with $\gamma=38.33\%$. The samples are sheared in the $x$-direction. The bar on the right gives the color code for the normalized densities shown in the snapshots, where the normalization is done with respect to the maximum value.}
    \label{fig:3dsnap}
\end{figure}

Finally, we performed an analogous comparison for the 3D experimental system. Representative snapshots of the 3D density maps are shown in Fig.~\ref{fig:3dsnap} and reveal again a pronounced degree of similarity among all quantities. 

To provide a quantitative assessment of the similarity between the local map of $D^2_{\text{min}}$ and the various defect densities introduced above, we consider the structural similarity index measure (SSIM) \cite{WangBovik2006},
\begin{equation}
\text{SSIM}[X,Y]\equiv \frac{\mathrm{Cov}(X,Y)}{\mathrm{Std}(X) \mathrm{Std}(Y)}.
    \label{eq:ssim}
\end{equation}

\begin{figure*}
\centering
\includegraphics[width=\linewidth]{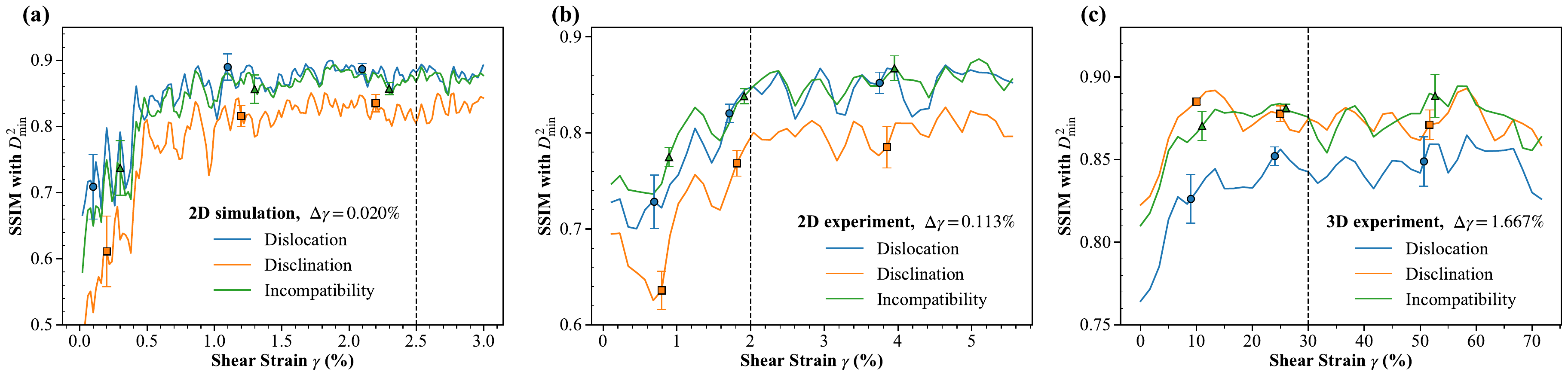}
\caption{\textbf{Benchmarking continuous density fields against $D^2_{\text{min}}$.} Structural similarity index measure (SSIM) as a function of strain for three different systems: \textbf{(a)} 2D simulation, \textbf{(b)} 2D experiment, and \textbf{(c)} 3D experiment. The solid lines represent averages over three independent samples, while the symbols show representative data points with corresponding error bars. Vertical dashed lines indicate the location of the yielding transition. 
}
    \label{fig:3}
\end{figure*}

The SSIM is obtained by mapping the fields onto a square/cubic lattice and calculating the covariance/variance of these mapped fields. Figure~\ref{fig:3} shows the SSIM between the various physical fields and $D^2_{\text{min}}$ as a function of shear strain for the three systems. 

For the 2D Lennard--Jones glass, Fig.\ref{fig:3}(a), the SSIM at very small strains is approximately $0.5$, $0.6$, and $0.7$ for the disclination, incompatibility, and dislocation densities, respectively. The correlation then increases rapidly and saturates at $\gamma^* \approx 1\%$, reaching values of about $0.82$, $0.86$, and $0.86$, further confirming the strong correspondence between these quantities and $D^2_{\text{min}}$. Importantly, $\gamma^*$ is significantly smaller than the yielding strain (vertical dashed line). This indicates that the high degree of similarity is not simply a consequence of large-scale plastic flow, but instead an intrinsic feature of these descriptors.

The weaker correlation at very small strain can be attributed to the dominance of elastic, nearly affine deformations, where both $D^2_{\text{min}}$ and the defect densities are weak and more susceptible to noise (see Appendix \ref{applow}).

Figure~\ref{fig:3}(b) presents the same similarity analysis for the 2D experimental data. Despite the very different nature of the interparticle potential and the presence of friction, the overall trend is consistent with the one of the simulations. We note that the curve for disclinations lies systematically below those for dislocations and incompatibility, suggesting that rotational defects play a subleading role for plasticity in two-dimensions. It will be interesting to perform a more extensive exploration of these findings to assess the universality of this statement beyond the two systems considered in this work.

Finally, we present in Fig.~\ref{fig:3}(c)  the analysis for the 3D granular system. Interestingly, we find that it is now the disclination density which exhibits the strongest correlation with $D^2_{\text{min}}$, indicating a stronger role of angular defects over translational ones in three dimensions. This is consistent with previous findings highlighting disclination-like defects as topological signatures of plasticity in 3D sheared granular matter \cite{Cao2018}, as well as the role of vortex-like structures in granular systems \cite{TORDESILLAS2016215}.

The observed distinction between two and three dimensions can be rationalized by the structure of the rotational phase space, further modulated by frictional interactions. In two dimensions, rotations are strictly confined to a single axis perpendicular to the plane (SO(2) symmetry), constraining plastic deformation to be predominantly mediated by translational slip modes. In three dimensions, the phase space for rotational motion expands to SO(3), allowing rotations about arbitrary axes and rendering rotational channels vastly more accessible for stress relaxation. While phase space geometry sets the fundamental capacity for these modes, frictional interactions act as a dynamic amplifier. Tangential forces convert macroscopic shear into local torques, driving particles along non-affine rotational trajectories and generating complex multi-axial dynamics that are not possible in 2D. Consequently, the synergy between an expanded rotational phase space and friction-induced torques promotes rotational defects to a dominant role in 3D plasticity.

The validity of our analysis is supported in Appendix \ref{appMSE}, where we show that an alternative similarity measure, the mean squared error (MSE), leads to consistent conclusions. Finally, in Appendices \ref{appstable} and \ref{appstrain}, we also examine the robustness of the correlations presented in Fig.~\ref{fig:3} under both increasing strain interval \(\Delta \gamma\) and the addition of external noise. These findings demonstrate the robustness of our conclusions under these changes of parameters.

Having benchmarked our continuous fields against the standard $D^2_{\text{min}}$ descriptor, we now take a step forward and analyze the cross-correlations among these density fields. This analysis is motivated by the visual spatial signatures in Figs.~\ref{fig:2} and \ref{fig:3dsnap}, which reveal striking structural similarities that warrant a more quantitative investigation.

\begin{figure*}[ht]
    \centering
    \includegraphics[width=\linewidth]{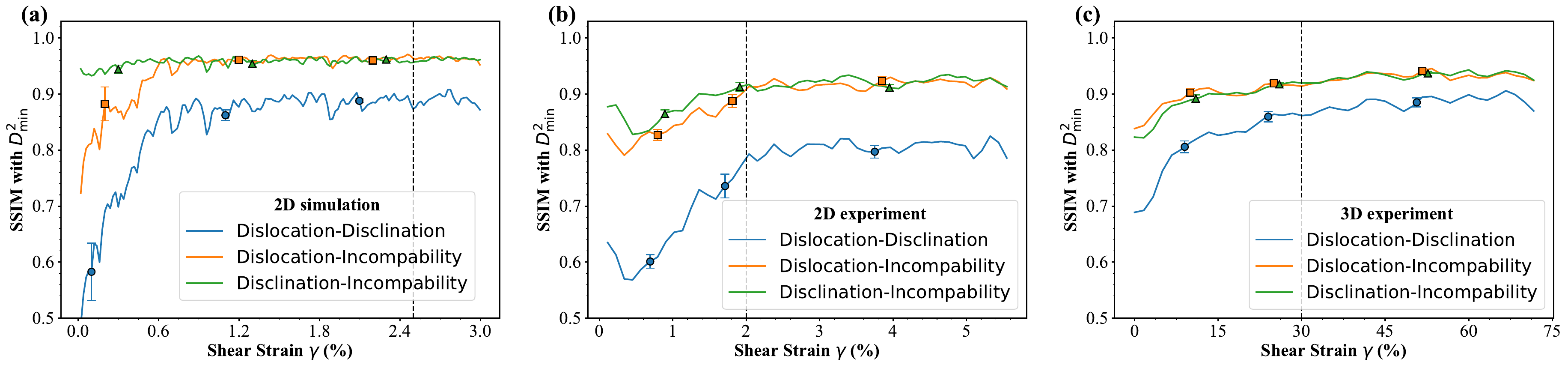}
    \caption{\textbf{Cross-correlations and spatial similarities among the density fields.} Structural Similarity Index Measure (SSIM) between the different density fields are shown as a function of strain for the \textbf{(a)} 2D simulation, \textbf{(b)} 2D experiment, and \textbf{(c)} 3D experiment. The vertical dashed line indicates the yielding strain.}
    \label{fig:ssim_all}
\end{figure*}

To quantify these visual impression of a correlation we make use of the SSIM introduced in Eq.~\eqref{eq:ssim} to measure the cross-correlation between the different density fields. 
The results for the three systems are reported in Fig.~\ref{fig:ssim_all}. For strains above yielding (vertical dashed line), both the dislocation--incompatibility and disclination--incompatibility correlations are very high, reaching values of approximately $90\%$. In contrast, within the quasi-elastic linear regime, the correlation between the disclination and incompatibility fields appears significantly stronger than the others. This indicates that at small strain the incompatibility is mainly of disclination nature, i.e., the particles start to rotate around each other, likely because in order to move, a particle will have to wedge into the gap formed between two neighboring particles.

Interestingly, at small strains, before defect clusters proliferate and percolate, the dislocation and disclination densities exhibit visibly different spatial structures. This demonstrates that, even in an amorphous system, these two fields probe distinct features of local plastic deformation. At larger strains below yielding, this distinction disappears as the SSIM increases to nearly $0.9$, making the two spatial patterns almost identical. Drawing intuition from classical defect mechanics, where a disclination dipole behaves as an edge dislocation, we are led to the speculation that a dense network of disclination-like structural defects naturally acts as an effective dislocation field.

\section{Outlook}
In this work, considering 2D and 3D simulated and experimental amorphous solids, we investigate the properties of density fields which in ordered materials reduce to standard indicators of plasticity -- dislocation, disclination, and incompatibility densities. We show that these defect fields, computed in amorphous systems from the displacement field under deformation, exhibit a striking similarity to the standard non-affine measure $D^2_{\text{min}}$, and thus emerge as effective descriptors of plasticity even in disordered materials. These observations are in line with the geometric framework recently proposed by Moshe \textit{et al.} \cite{doi:10.1073/pnas.1506531112,PhysRevE.107.055004,PhysRevE.105.L043001,Kumar_2024} and the picture of \textit{continuous defects} in the sense of Kl\'eman \cite{RevModPhys.80.61}.

This connection permits to have a unified description of plasticity across ordered and disordered materials: The fields that localize the discrete defects in crystals become in amorphous systems smooth quantities that signal the presence of enhanced plasticity. These fields are directly linked to the density of topological defects or the strength of incompatibility, making them valuable for describing plasticity even in the absence of well-defined underlying singularities—unlike other approaches based on vortices, e.g., \cite{wu2023topology,Vaibhav2025,bera2025hedgehogtopologicaldefects3d,wu2024geometrytopologicaldefectsglasses,bedollamontiel2025strikingsimilaritiesdynamicsvibrations,10.1093/pnasnexus/pgae315,PhysRevE.109.L053002,PhysRevB.110.014107,wang2025topological,bera2025microscopic}. The fact that these fields take continuous values makes them an attractive alternative to approaches that consider only discrete defects or that require some sort of cut-off to define them. Note that due to their intrinsic disorder, glasses are not structures that can be expected to have defects that are well defined in a mathematical sense~\cite{baggioli2026topological}, but instead to have zones in which plasticity is more/less likely. Hence the fields presented here allow to take into account this important property of glasses in a natural way.

In addition, the ability of these fields to distinguish between rotational and translational defects is particularly valuable, as it provides deeper insight into the nature of plastic events---something $D^2_{\min}$ cannot capture. This capability has already revealed a key difference between plasticity in two and three dimensions, where the roles of translational and rotational defects are reversed.

We also note that applying the same methods to the eigenvector fields of vibrational modes in the unperturbed configuration has been found to provide a robust predictor of plastic activity under deformation~\cite{newNye}, i.e., this method is not only applicable as a purely \textit{a posteriori} characterization. 

Finally, from a theoretical perspective, it will be interesting to establish a more direct connection between these densities and the geometric charges introduced by Moshe \textit{et al.} \cite{doi:10.1073/pnas.1506531112,PhysRevE.107.055004,PhysRevE.105.L043001,Kumar_2024}, since such a link could provide a better understanding of how defects relate to the anomalous elastic behavior observed in amorphous materials. This point is also related to the question whether our density fields can be understood as a continuous distribution of Eshelby-like defects. We believe this is indeed the case and leave a detailed exploration of this idea for the near future.

\section*{Acknowledgments} 
We thank M.~Falk, P.~Harrowell, M.~Moshe, G.~Szamel, C.~Du, A.~Zaccone, I.~Regev, Z.~Wu, J.-L. Barrat, A.~Bera, L.~Huang, Y.~Wang, and A.~Liu for useful discussions on this topic. MB acknowledges the support of the Shanghai Municipal Science and Technology Major Project (Grant No.2019SHZDZX01) and the support of the sponsorship from the Yangyang Development Fund. JS and JZ acknowledge the support of the NSFC (No.11974238, No.12274291 and No.12534008) and the Shanghai Municipal Education Commission Innovation Program under No. 2021-01-07-00-02-E00138. JS and JZ also acknowledge the support from the SJTU Student Innovation Center. XW, YX and YW are supported by the National Natural Science Foundation of China (No. 12274292).

\appendix

\counterwithin{figure}{section}
\renewcommand{\thefigure}{\thesection\arabic{figure}}

\section{Ideal topological defects and their singular contribution to the continuous defect densities}\label{idealsec}

Given the particle trajectories, one can construct a continuous displacement field 
$u_i(\mathbf{x})$, where $\mathbf{x}=(x,y,z)$ in 3D and $\mathbf{x}=(x,y)$ in 2D. 
The Burgers vector is then defined as the circulation 
of the displacement field along a closed loop:
\begin{equation}
    b_j(\mathbf{x}) = \oint d u_j = \oint \partial_k u_j(\mathbf{x}) \, dx_k .
\end{equation}
Applying Stokes' theorem and using the Einstein summation convention, this can be rewritten as a surface integral:
\begin{equation}
    b_j(\mathbf{x}) = \int \epsilon_{ilk}\, \partial_l \partial_k u_j(\mathbf{x}) \, dS_i,
\end{equation}
which naturally leads to the definition of the Nye tensor,
\begin{equation}
    \alpha_{ij}(\mathbf{x}) = \epsilon_{ilk}\, \partial_l \partial_k u_j(\mathbf{x}),
\end{equation}
describing, in crystals, the local density of dislocations.

For a smooth (single-valued and differentiable) displacement field, mixed partial 
derivatives commute, and therefore $\alpha_{ij}(\mathbf{x})=0$. 

In two dimensions, the only nonzero components of the Nye tensor are
\begin{equation}
    \alpha_{z i} = \epsilon_{zmn}\, \partial_m \partial_n u_i 
    = (\partial_x \partial_y - \partial_y \partial_x) u_i 
    = [\partial_x,\partial_y]\, u_i,
\end{equation}
which explicitly measures the non-commutativity of derivatives and thus the 
local defect content.

Let us consider the canonical displacement field of a 2D dislocation with Burgers vector along $x$,
\begin{equation}
    u_x = \frac{b}{2\pi} \arctan\!\left(\frac{y}{x}\right) + u_x^{\mathrm{reg}}, 
    \qquad 
    u_y = u_y^{\mathrm{reg}},
\end{equation}
where $u^{\mathrm{reg}}$ denotes a smooth (regular) contribution. Since the Nye tensor 
only probes singular terms, the smooth part does not contribute and can be ignored.

It follows immediately that
\begin{equation}
    \alpha_{zy} = [\partial_x,\partial_y] u_y = 0,
\end{equation}
and we only need to analyze the singular part of $u_x$.

Away from the origin, $(x,y)\neq(0,0)$, derivatives commute and one finds
\begin{equation}
    \partial_x u_x = -\frac{b}{2\pi} \frac{y}{x^2 + y^2}.
\end{equation}
However, the function $\arctan(y/x)$ is multivalued and exhibits a branch cut 
(along, e.g., the negative $x$-axis), which generates a singular contribution. 
Taking this into account in the sense of distributions, one obtains
\begin{equation}
    [\partial_x,\partial_y]\, u_x 
    = \partial_x \partial_y u_x - \partial_y \partial_x u_x 
    = b\,\delta(x)\delta(y).
\end{equation}

Therefore,
\begin{equation}
    \alpha_{zx}(\mathbf{x}) = b\,\delta(x)\delta(y),
\end{equation}
showing that the Nye tensor is localized at a point and corresponds to a 
topological defect with Burgers vector magnitude $b$.

To consider the case of a disclination, we introduce the local rotation (vorticity) as
\begin{equation}
    \omega_i(\mathbf{x})  = \frac{1}{2}\,\epsilon_{ijk}\,\partial_j u_k(\mathbf{x}) .
\end{equation}
The antisymmetric quantity $\omega_i(\mathbf{x}) $ captures local rigid-body rotations.

In direct analogy with the Burgers vector for dislocations, one can define 
defect densities associated with the rotation field. These are described by 
the Frank tensor, which quantifies in crystals the density of disclinations.

A simple analytical displacement field describing a combined dislocation 
(with Burgers vector $\mathbf{b}$) and disclination (with strength $d_{xy}$) is
\begin{equation}
\begin{aligned}
    u_x(\mathbf{x}) &= \frac{b_x + d_{xy} y}{2 \pi} \arctan\!\left(\frac{y}{x}\right)
    + \frac{-b_y - d_{xy} x}{4 \pi} \ln(x^2 + y^2), \\
    u_y(\mathbf{x}) &= \frac{b_y - d_{xy} x}{2 \pi} \arctan\!\left(\frac{y}{x}\right)
    + \frac{b_x - d_{xy} y}{4 \pi} \ln(x^2 + y^2).
\end{aligned}
\end{equation}

From this, the scalar vorticity (in 2D) reads
\begin{align}
    \omega(\mathbf{x}) &= \frac{1}{2}(\partial_x u_y - \partial_y u_x)\nonumber
    \\&= -\frac{d_{xy}}{2 \pi} \arctan\!\left(\frac{y}{x}\right)
    + \frac{1}{4 \pi} \frac{y b_y - x b_x}{x^2 + y^2}.
\end{align}

Both the displacement field and the rotation contain the multivalued function 
$\arctan(y/x)$, which introduces a branch cut. As a consequence, derivatives 
do not commute in the distributional sense, leading to nonzero defect densities. 
Specifically:
\begin{itemize}
    \item The non-commutativity of derivatives acting on $u_i(\mathbf{x})$ gives a nonzero 
    Nye tensor, corresponding to a dislocation localized at the origin.
    \item The non-commutativity of derivatives acting on $\omega(\mathbf{x})$ gives a nonzero 
    Frank tensor, corresponding to a disclination.
\end{itemize}

\section{Computational methods}\label{appcomp}

Since the particle displacements are associated with the positions of the particle and the latter are distributed in space in a disordered manner, this field is defined only at discrete points. To compute spatial derivatives one needs, however, a continuous displacement field. Since derivatives cannot be evaluated directly on irregularly spaced data points, we map the displacement field onto a regular grid and then compute derivatives using central finite differences.

\begin{figure}[ht]
    \centering
    \includegraphics[width=1\linewidth]{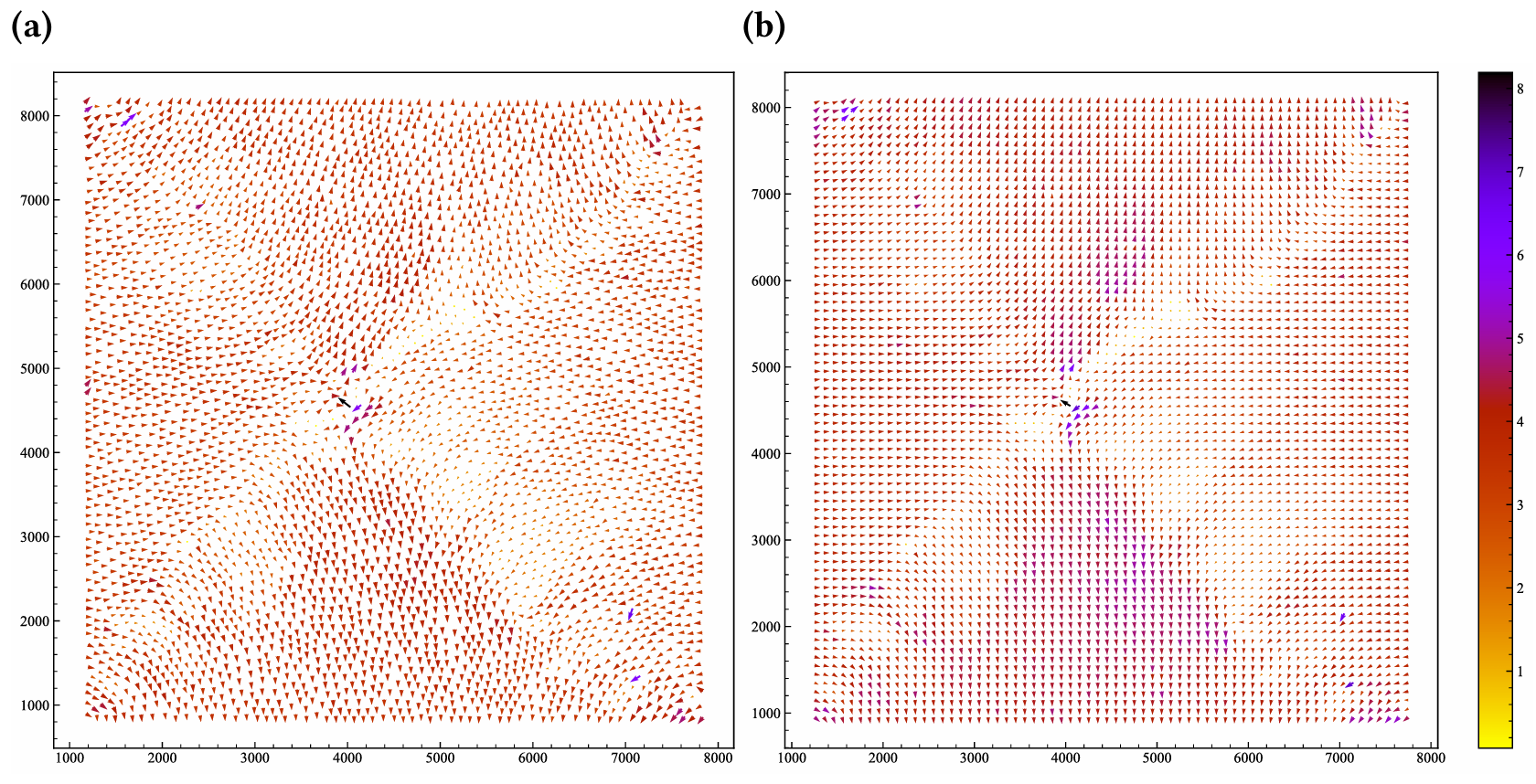}
    \caption{Two-dimensional experimental system: \textbf{(a)} Displacement vectors defined at particle positions. \textbf{(b)} Continuous displacement field obtained by interpolation onto a grid with spacing $a = 1.0 \,d_s$ with $d_s$ the diameter of the small particles.
    }\label{fig:S1}
\end{figure}

First, we apply linear interpolation to obtain displacement values on a grid with spacing $a$. Thus for the 2D systems we have
\begin{equation}
u(x, y) \longrightarrow u(i, j) \,,
\end{equation}

\noindent
where $i$ and $j$ give the position on the lattice. The displacement field on the lattice has the same information as the particle displacement. As demonstrated in Fig.~\ref{fig:S1}, the main features of the different fields do not depend in a significant manner on the interpolation procedure, i.e., this approach allows to make a robust analysis of their characteristics.

We then approximate the spatial derivatives by central differences:
\begin{align}
\pdv{u}{x}(x,y)
&= \lim_{\Delta x \to 0}
\frac{u(x + \tfrac{1}{2}\Delta x, y) - u(x - \tfrac{1}{2}\Delta x, y)}{\Delta x}
\\[4pt]
\pdv{u}{y}(x,y)
&= \lim_{\Delta y \to 0}
\frac{u(x, y + \tfrac{1}{2}\Delta y) - u(x, y - \tfrac{1}{2}\Delta y)}{\Delta y},
\end{align}
which gives for the discrete fields
\begin{align}
\pdv{u}{x}(i,j)
= \frac{u(i+1,j) - u(i-1,j)}{2},
\\[4pt]
\pdv{u}{y}(i,j)
= \frac{u(i,j+1) - u(i,j-1)}{2}.
\end{align}

Higher order derivatives are obtained in the same manner. Finally, we apply a Gaussian smoothing with width of $1.5\,d$ (with $d$ the particle diameter) to reduce numerical noise in the resulting fields for all systems.

\section{Defects captured by continuous densities and comparison to vortex-like defects}\label{appB}
\begin{figure}[ht]
    \centering
    \includegraphics[width=1\linewidth]{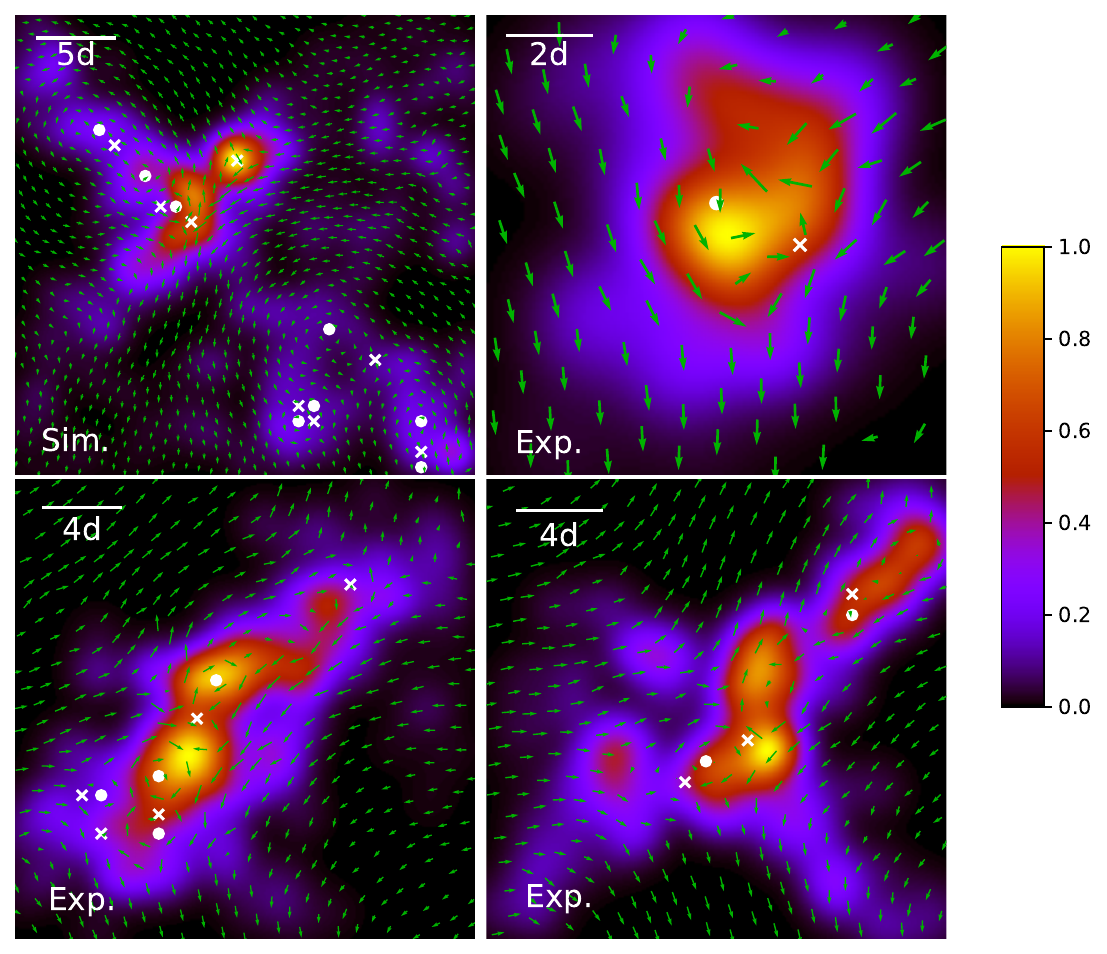}
    \caption{Snapshots of the two-dimensional particle displacement field, highlighting the geometric nature of the defects captured by the continuous dislocation density (yellow/orange areas). The dislocation density is normalized to its maximum value in each panel, with the color scale bar shown on the right. The top-left panel corresponds to the 2D simulated Lennard-Jones glass, while all other panels are for the 2D experimental granular system. White crosses and  disks denote vortex-like defects with winding numbers $w=-1$ (anti-vortices) and $w=+1$ (vortices), respectively. In all panels, $d$ indicates the particle diameter. Sim. and Exp. refer to the 2D simulation and experimental data, respectively.
    }\label{tdsnapshot}
\end{figure}

To clarify the nature of the geometric defects captured by the continuous densities introduced in this work, we present in Fig.~\ref{tdsnapshot} additional snapshots of the particle displacement field for both the simulated 2D Lennard-Jones system and the experimental 2D granular system. Yellow–orange regions indicate the location of plastic defects, i.e., localized areas where the Nye dislocation density attains large values.

For completeness, we also consider topological defects in the phase angle of the displacement field, defined through the winding number
\begin{equation}
    w \equiv \frac{1}{2\pi} \oint d\theta \,,
\end{equation}
where \(\theta\) denotes the local orientation of the displacement vector field. We evaluate \(w\) over the smallest available loops within the regular grid and identify regions with \(w=+1\) as topological vortices and those with \(w=-1\) as anti-vortices.

In Fig.~\ref{tdsnapshot}, these phase defects are superimposed on the continuous Nye dislocation density for the 2D systems. Anti-vortices and vortices are marked by white crosses and white disks, respectively. We observe a good correlation between regions of high dislocation density and areas with an elevated local density of vortex-like defects. Notably, all regions with large dislocation density involve at least one vortex dipole, and often more complex configurations composed of defects with alternating sign. In contrast, isolated vortex-like defects with $+1$ winding number rarely produce a noticeable signature in the dislocation density. This suggests that it is the collective accumulation of such defects—rather than their existence as individual entities—that is essential for capturing plastic rearrangements in glasses. Moreover, it highlights the greater significance of $-1$ defects,  $+/-$ dipoles, and more complex clusters compared to isolated $+1$ vortices.

\section{Demonstration of the robustness of plastic defect structures with respect to grid size, interpolation scheme, and grid positioning}\label{apps1}

To validate the robustness of our calculations of the density fields, we systematically varied the interpolation grid size, the interpolation scheme, and the relative positioning of the grid used to compute the various fields in the simulated system.

\begin{figure}[ht]
\centering
\includegraphics[width=1\linewidth]{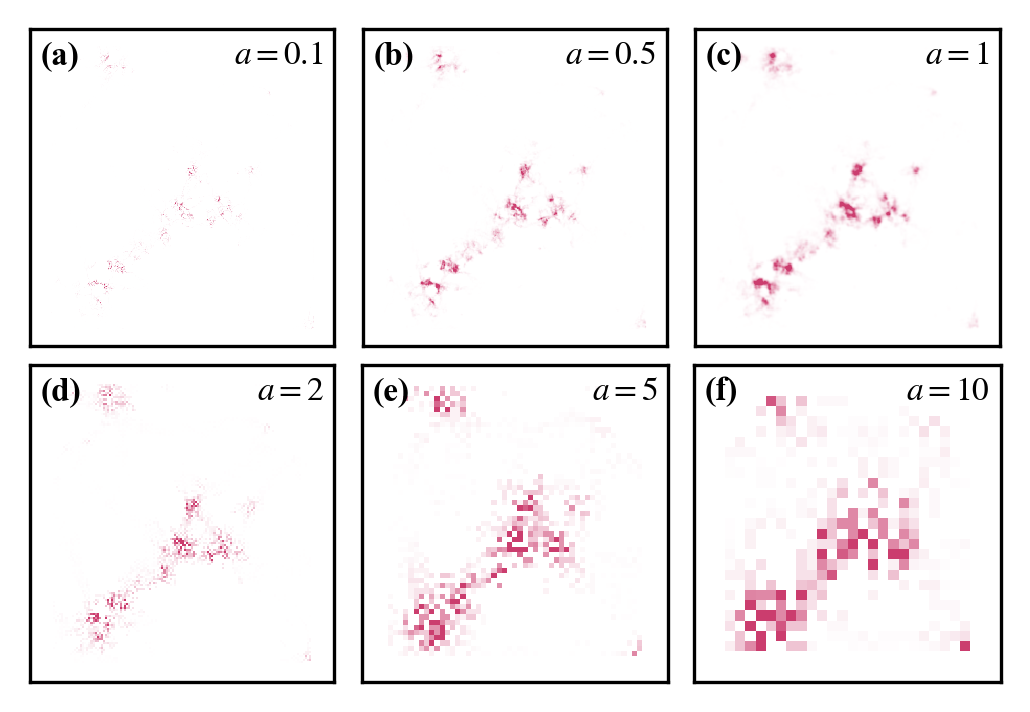}
\caption{Dislocation density amplitude $|\vec{\alpha}(x,y)|$ computed using different grid sizes at $\gamma=3.4\%$ for the 2D simulation. Panels \textbf{(a–f)} correspond to grid sizes ranging from $0.1$ to $10.0$, as indicated in each subplot. The color intensity  represents the local magnitude of the dislocation density. All fields are normalized to have values in an interval $[0,1]$.
}
\label{in_grid}
\end{figure}

By varying the grid size (Fig.~\ref{in_grid}), we find that the locations of high density regions remain unchanged across a wide range of resolutions. This demonstrates that our results are not sensitive to the discretization scale, indicating that these regions correspond to intrinsic regions of anomalously strong spatial variation in the underlying deformation field, whose identification is independent of the specific interpolation grid.

\begin{figure}[ht]
\centering
\includegraphics[width=1\linewidth]{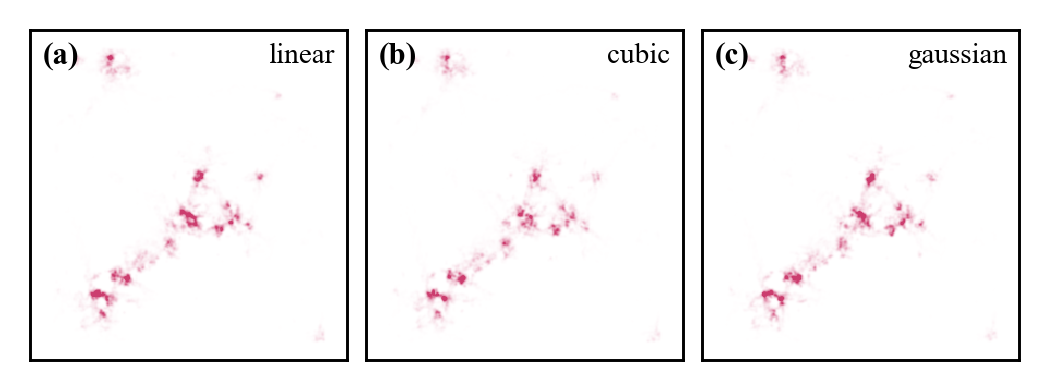}
\caption{Dislocation density for the 2D simulated system obtained using \textbf{(a)} linear, \textbf{(b)} cubic, and \textbf{(c)} Gaussian interpolation of the displacement field. Here, $\gamma=3.4\%$. The color intensity represents the local magnitude of the dislocation density.}
\label{in_method}
\end{figure}

Furthermore we tested whether the interpolation scheme used to calculate the derivatives of the displacement field affects the results. For this we considered three schemes—linear, cubic, and Gaussian and present the results in Fig.~\ref{in_method}. The resulting patterns are nearly identical, indicating that the observed singular structures are not artifacts of the interpolation procedure.

\begin{figure}[ht]
\centering
\includegraphics[width=1\linewidth]{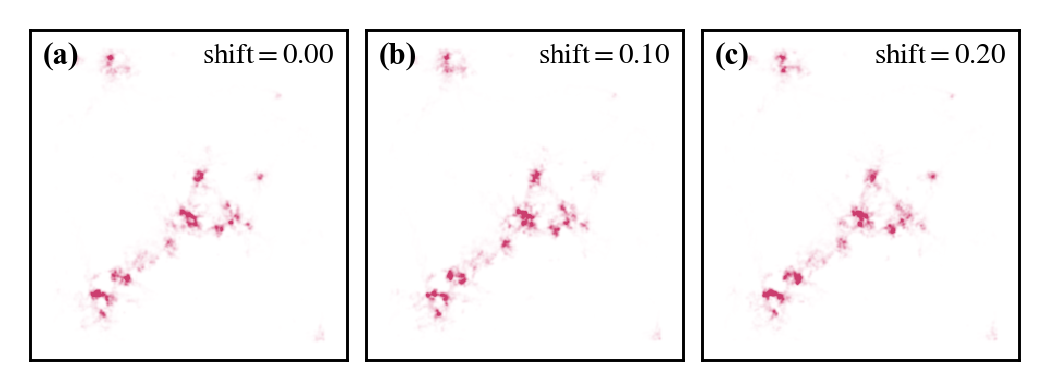}
\caption{Effect of shifting the interpolation grid relative to the system boundary in the 2D simulated system. Here, $\gamma=3.4\%$. The initial offset is $5.0$ with additional shifts along the diagonal direction of \textbf{(a)} $0.0$, \textbf{(b)} $0.1$, and \textbf{(c)} $0.2$. The color intensity represents the magnitude of the dislocation density.}
\label{in_shift}
\end{figure}

Finally, we examined the effect of shifting the grid relative to the system boundaries (Fig.~\ref{in_shift}). Specifically, we changed the initial and final positions of the interpolation, thereby effectively shifting the positions of the interpolation grid for different shift values. By excluding the boundary regions and varying the grid offset, we again observe that the locations of high topological density remain invariant. This further confirms that the identified features are intrinsic to the displacement field and not introduced by the interpolation or the alignment of the grid.

\section{Robustness of the dislocation density under external artificial noise}\label{apps2}
\begin{figure*}[ht]
    \centering
    \includegraphics[width=\linewidth]{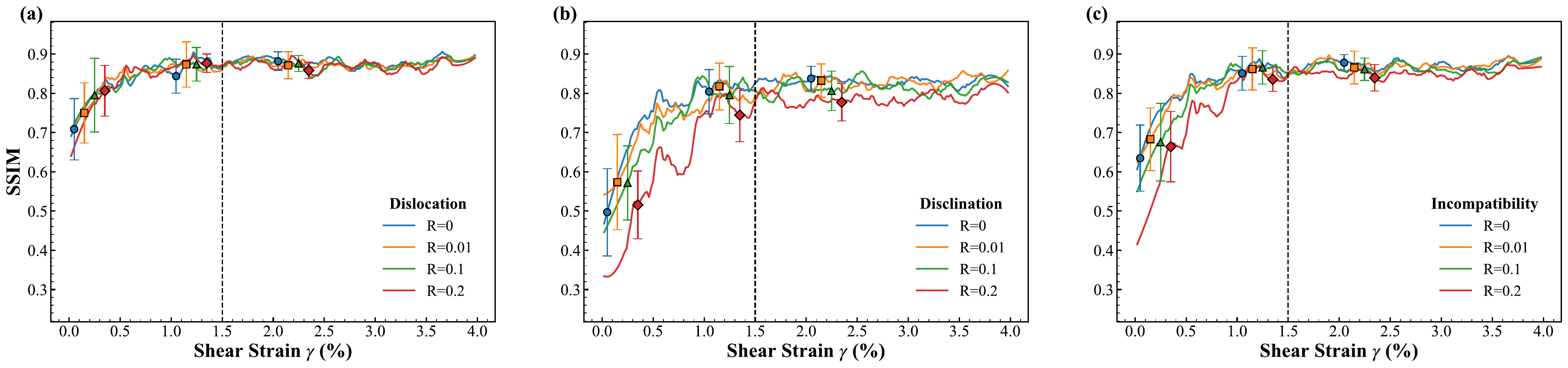}
    \caption{Structural Similarity Index Measure (SSIM) for 
    increasing artificial noise in the displacement field (see Eq.~\eqref{eq:noise}) for the 2D Lennard-Jones simulation. Panels (a)--(c) show, respectively, the SSIM between the dislocation, disclination, and incompatibility densities and $D^2_{\text{min}}$. The vertical dashed line indicates the yielding strain of the simulated system. }
    \label{fig:ssim_noise}
\end{figure*}

It is important to verify to what extent the various vector fields studied in the present work are affected by noise, since thermal fluctuations and experimental errors will prevent to have a perfect knowledge of these fields. To study these effects, we have added artificial noise to the displacement field:
\begin{equation}
    v_{\text{noise},i}(\vec{x}) = v_i(\vec{x})\left[1 + R\,\xi_i(\vec{x})\right], \qquad i=x,y,z,
    \label{eq:noise}
\end{equation}
where each component of the random vector $\vec{\xi}$ follows a Gaussian distribution with zero mean and unit standard deviation, and $R$ is the amplitude of the noise. The influence of this noise on the structural similarity index measure (SSIM) for the case of the 2D Lennard-Jones system is presented in Fig.~\ref{fig:ssim_noise}, for different noise amplitudes $R$.

As the noise amplitude increases, the similarity between the disclination and incompatibility densities and $D^2_{\min}$ decreases, particularly in the pre-yield regime ($\gamma < 1\%$). On the other hand, the dislocation density consistently retains a higher similarity compared to the others. This indicates that the dislocation density exhibits superior robustness against external perturbations. 

\begin{figure*}[ht]
    \centering
    \includegraphics[width=\linewidth]{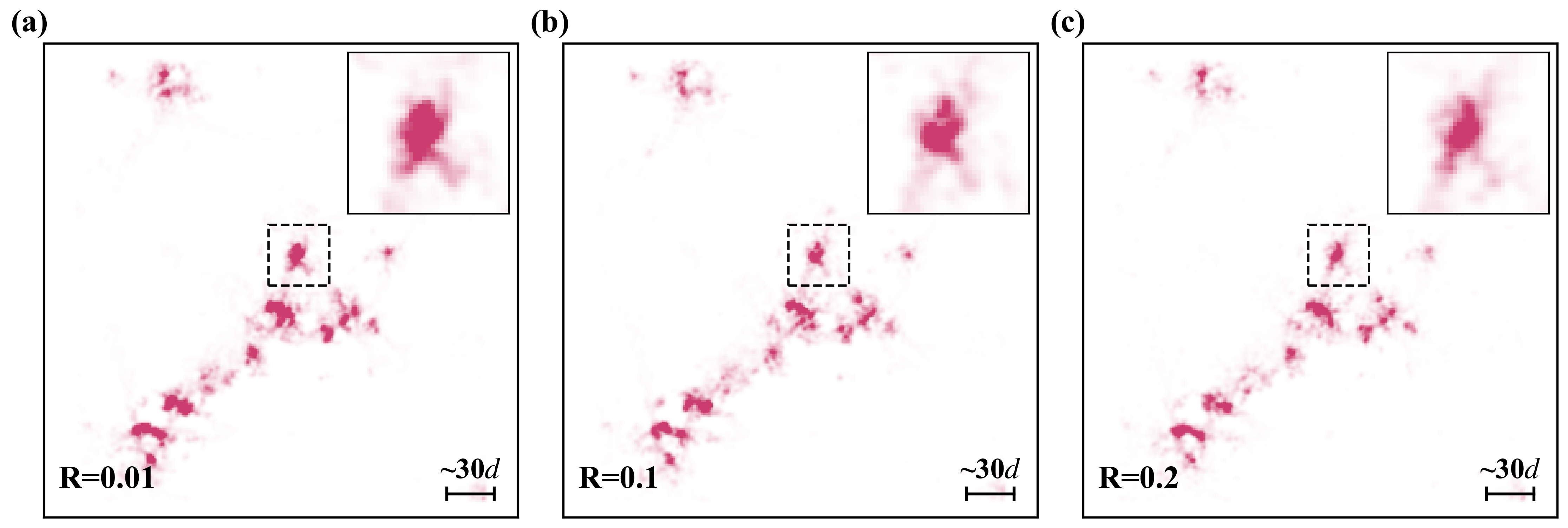}
    \caption{Snapshots of the dislocation density for the 2D simulation at $\gamma = 3.84\%$ with added noise levels of $0.01$, $0.1$, and $0.2$, corresponding to panels \textbf{(a)–(c)}, respectively. In each panel, the inset displays a magnified view of the dashed region in the main plot. 
    }
    \label{fig:noise_snap}
\end{figure*}

In Fig.~\ref{fig:noise_snap}, we present the spatial structure of the dislocation density for various levels of noise. One recognizes that this structure is indeed not affected significantly when the noise level is increased.

\begin{figure*} [ht]
    \centering
    \includegraphics[width=\linewidth]{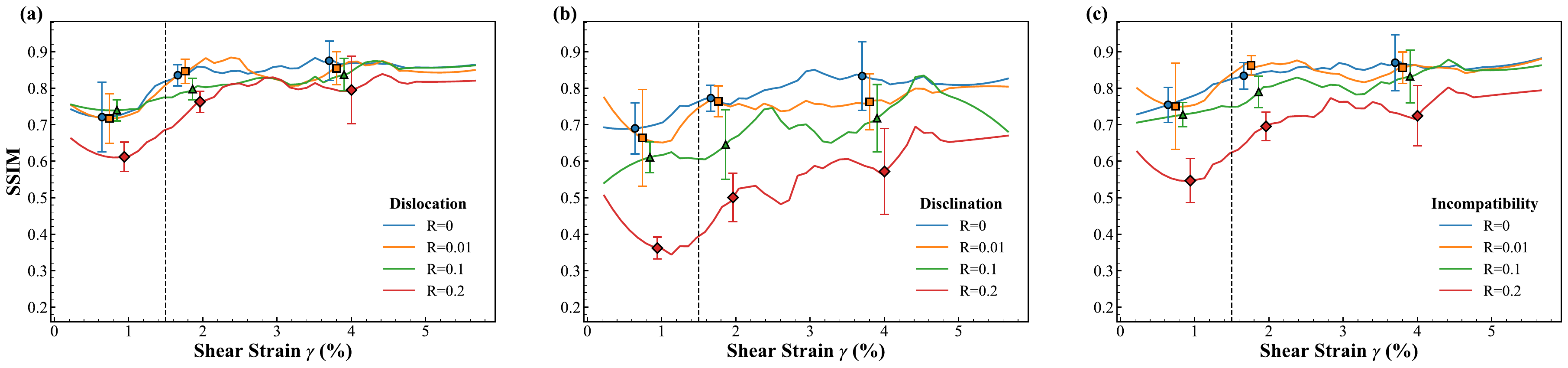}
    \caption{Structural Similarity Index Measure (SSIM) for  increasing artificial noise in the displacement field (see Eq.~\eqref{eq:noise}) for the 2D experimental data. Panels (a)--(c) show, respectively, the SSIM between the dislocation, disclination, and incompatibility densities and $D^2_{\text{min}}$. The vertical dashed line indicates the yielding strain of the experimental system.
    }
    \label{fig:noise_exp2d}
\end{figure*}

\begin{figure*} [ht]
    \centering
    \includegraphics[width=\linewidth]{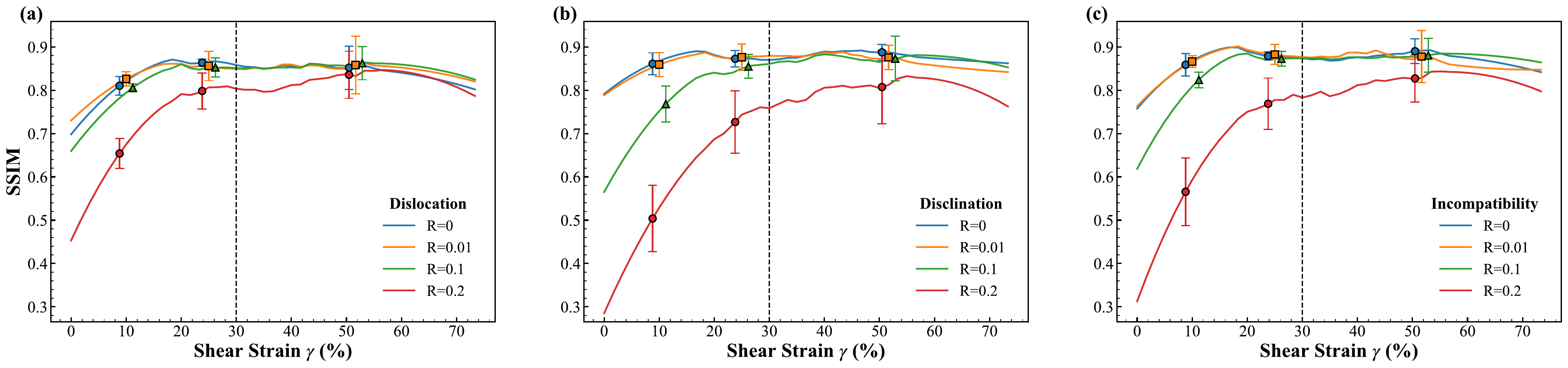}
    \caption{Structural Similarity Index Measure (SSIM) for increasing artificial noise in the displacement field (see Eq.~\eqref{eq:noise}) for the 3D experimental data. Panels (a)--(c) show, respectively, the SSIM between the dislocation, disclination, and incompatibility densities and $D^2_{\text{min}}$. The vertical dashed line indicates the yielding strain of the experimental system.
    }
    \label{fig:noise_exp3d}
\end{figure*}

We have repeated the same type of perturbation for the  2D experimental and 3D experimental systems and in Fig.~\ref{fig:noise_exp2d} and ~\ref{fig:noise_exp3d}, we present the corresponding results. As in the simulations, the dislocation density remains a robust indicator except at the largest noise level ($R=0.2$), whereas the other two densities are more strongly affected by artificial noise, showing a noticeable drop in SSIM with $D^2_{\min}$. This confirms the superior stability of the dislocation density as a measure of plasticity in for all systems.

\section{Extended analysis on continuous defect densities in the 2D experimental granular systems}\label{appmore}

\begin{figure*}[ht]
    \centering
    \includegraphics[width=1\linewidth]{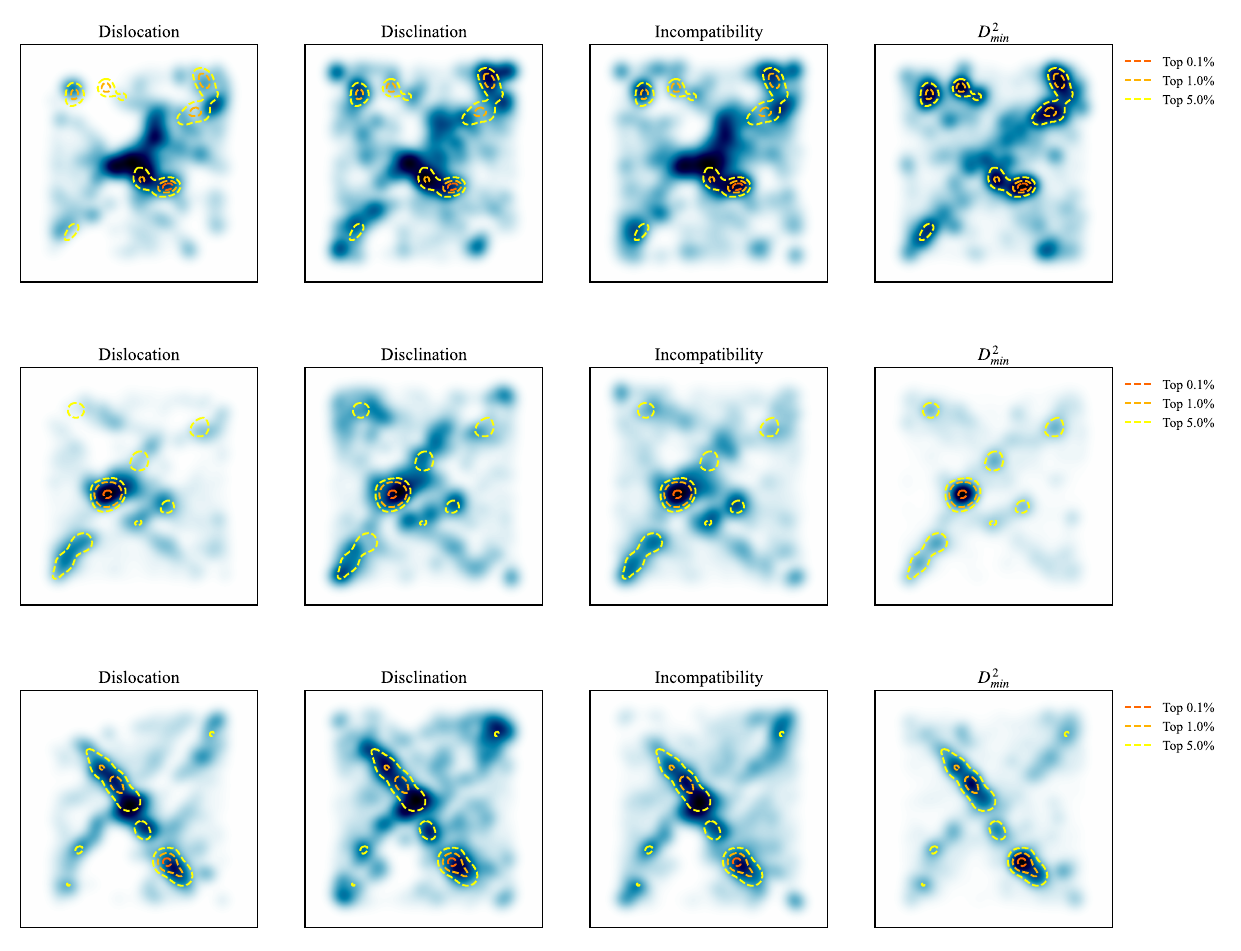}
    \caption{Heat maps of the dislocation, disclination, and incompatibility densities, together with $D^{2}_{\text{min}}$, for the 2D experimental data. For the dislocation density, we plot the norm of $\vec{\alpha}$. The applied strains are (top row) $\gamma = 2.83\%$, (middle row) $\gamma = 3.40\%$, and (bottom row) $\gamma = 3.96\%$.  }
    \label{fig:S2}
\end{figure*}

In Fig.~\ref{fig:S2}, we provide further examples in which we compare the fields for the dislocation, disclination and incompatibility densities, and $D^2_{\text{min}}$ for different values of the shear strain $\gamma$. The corresponding similarity degree is shown in Fig.~\ref{fig:3}(b). For all cases one finds a good correspondence between the spatial structure of the different fields. 

\section{Effect of visualization scale on apparent blur in Figure 3 in main text}\label{appblur}
\begin{figure}[h]
    \centering
    \includegraphics[width=\linewidth]{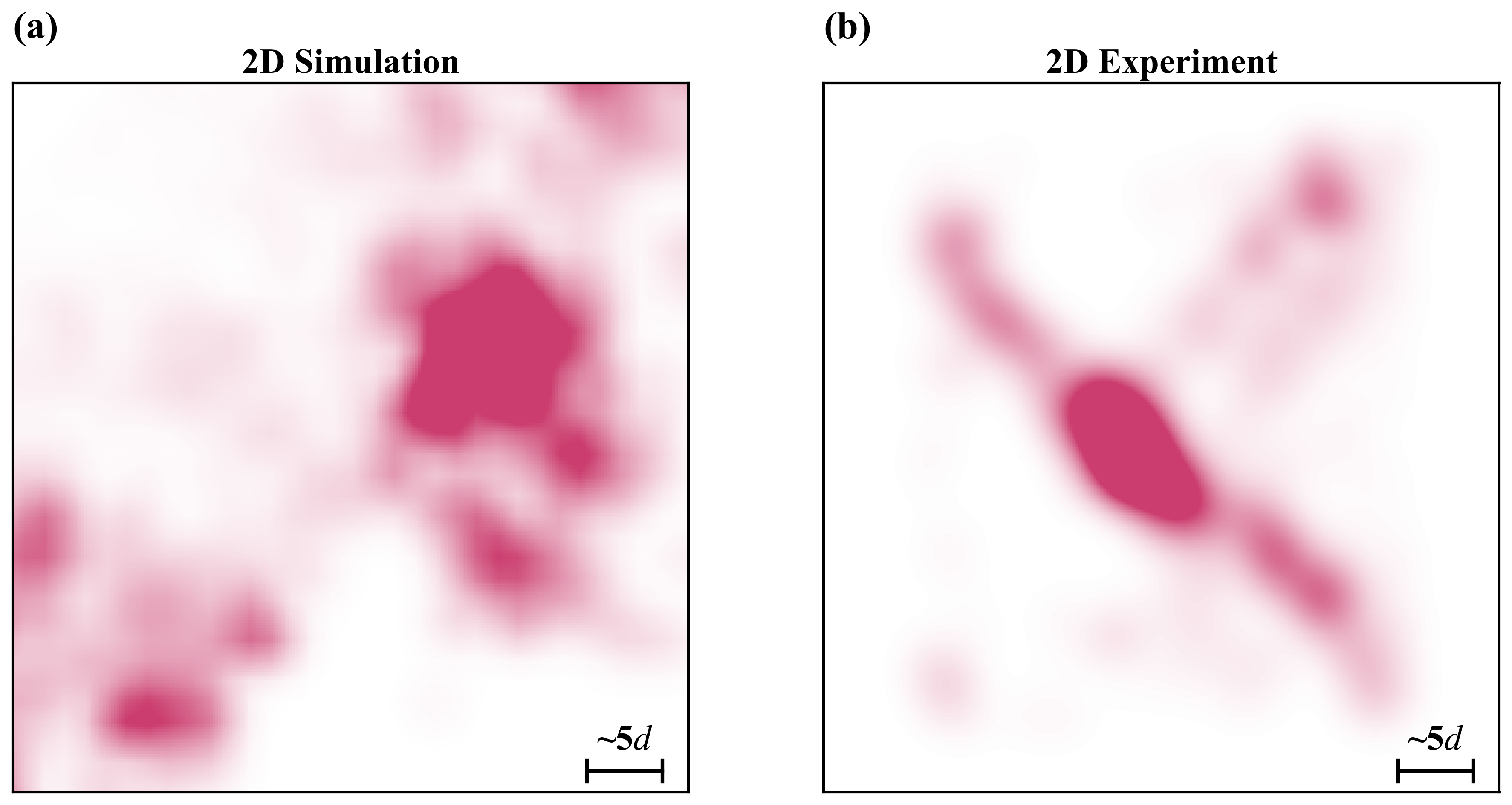}
    \caption{\textbf{(a)} Magnified view of the defect density maps for the 2D simulations shown in Fig.~\ref{fig:2}(a) of the main text (enlarged by a factor of $30/4$). This rescaling places the sample, in units of the particle diameter, on a length scale comparable to that of the 2D experimental system (see scale bar). \textbf{(b)} Experimental 2D data displayed at the same length scale. The simulation and experimental images exhibit a comparable level of blur in the density field.}
    \label{fig:zoom_in}
\end{figure}

In Fig.~\ref{fig:2} in the main text, we noted that the experimental data appear more blurred than the simulation results. We attributed this difference to the disparity in sample size. To verify this, we rescaled and cropped the simulation data to match the magnification and image dimensions of the experimental data, see Fig.~\ref{fig:zoom_in}. The resulting images display a comparable level of blur, indicating that the difference in sharpness arises from visualization effects rather than any underlying physical difference between the experimental and simulated systems.

\section{Continuous defect densities in the 2D Lennard-Jones simulated glass at low strain}\label{applow}
\begin{figure*}[ht]
    \centering
    \includegraphics[width=1.0\linewidth]{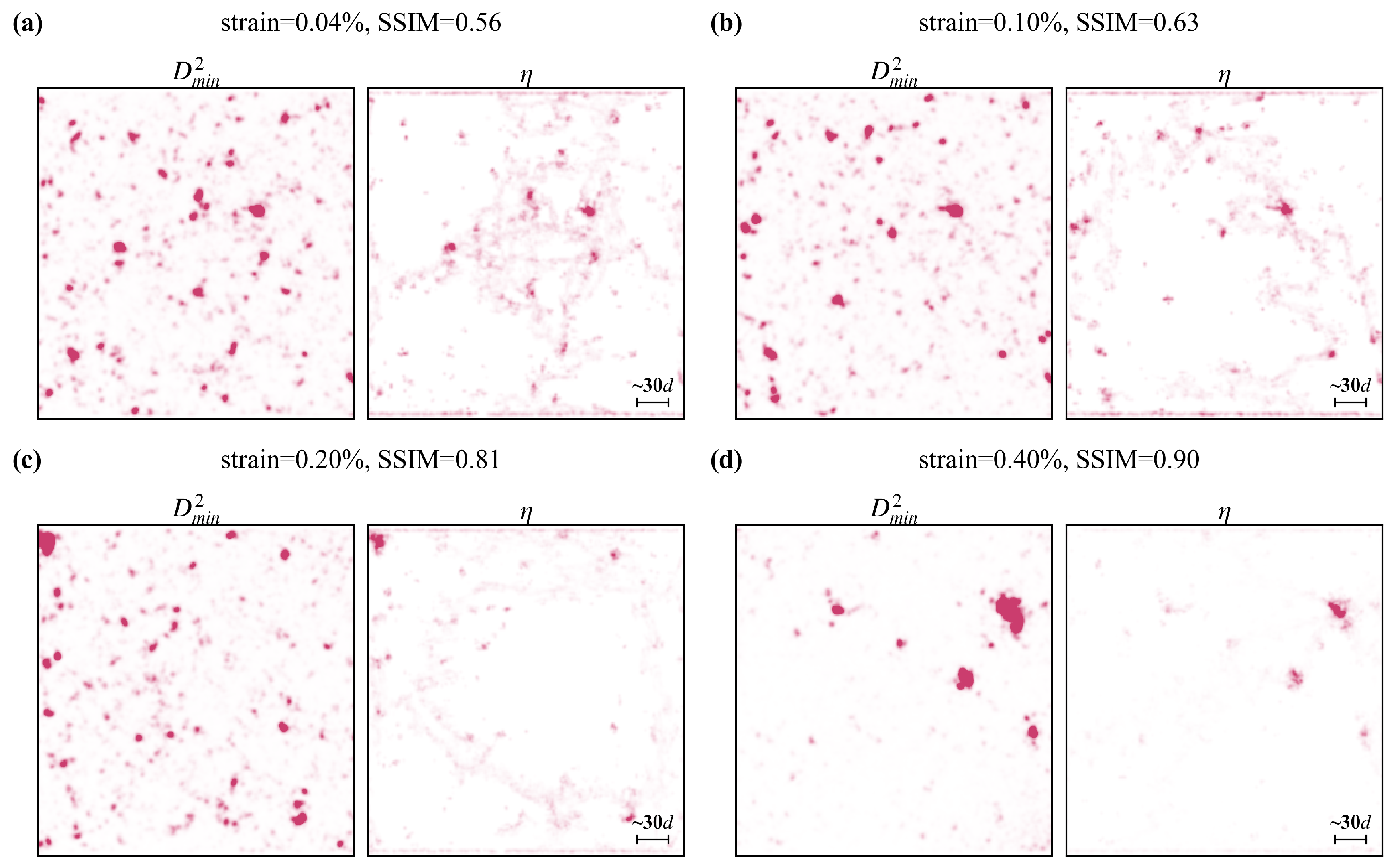}
    \caption{Comparison between $D^2_{\text{min}}$ and incompatibility density $\eta$ at different low strain for the 2D Lennard-Jones glass, see headings of panels for the values of $\gamma$. The corresponding SSIM values are reported on top of each panel.
    }
    \label{low_strain}
\end{figure*}

In the main text (Fig.~\ref{fig:2}) we presented a direct comparison between the defect densities and $D^2_{\text{min}}$ at a relatively large strain, $\gamma = 3.38\%$. 
For completeness we present in Fig.~\ref{low_strain} four additional comparisons between $D^2_{\text{min}}$ and the incompatibility density $\eta$ at low strain. Panels (a)–(c) correspond to $\gamma \lesssim 0.2\%$, where the SSIM with $D^2_{\text{min}}$ (see Fig.~\ref{fig:3} in the main text) remains moderate (in the range $50\%$–$80\%$). In this regime, the response is predominantly elastic, with only sparse plastic rearrangements of small amplitude. The non-affine measure $D^2_{\text{min}}$ is noisy and spatially delocalized. While a clear visual similarity between $\eta$ and $D^2_{\text{min}}$ is already evident, the quantitative agreement is limited by the substantial background noise.

By contrast, panel (d) shows a slightly larger—yet still small—strain ($\gamma=0.4\%$), where the SSIM increases substantially. In this case, plastic events become fewer but more pronounced, leading to a much clearer correspondence between the two measures.

A step-by-step comparison is provided in the Supplementary Video \texttt{d2min\_density.mp4} .

\section{Mean Squared Error similarity analysis} \label{appMSE}
\begin{figure*}[h]
    \centering
    \includegraphics[width=\linewidth]{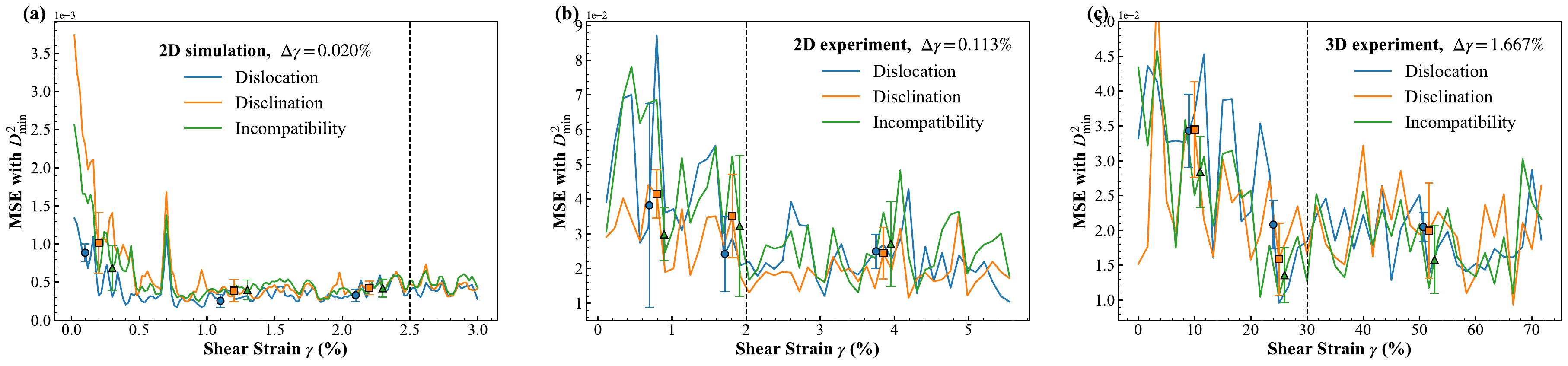}
    \caption{Mean Squared Error (MSE) as a function of strain for \textbf{(a)} 2D simulation, \textbf{(b)} 2D experiment, and \textbf{(c)} 3D experiment.}
    \label{fig:mse}
\end{figure*}

In the main text we have used the SSIM to quantify the similarity between the various fields. Another possibility to do this is via the Mean Squared Error (MSE) analysis. Given two fields $X_i$ and $Y_i$ whose values are normalized between $0$ and $1$, where $i$ indicates the position on the lattice, the MSE is defined as
\begin{align}
    \text{MSE} = \frac{1}{n} \sum_{i=1}^{n} (X_i - Y_i)^2 ,
\end{align}
where $n$ is the number of grid points. A small MSE value indicates thus a good match between the two fields. Figure~\ref{fig:mse} shows the MSE between the various density fields and $D^2_{\text{min}}$. In all cases, the MSE remains very small, confirming the high degree of similarity between these quantities reported in the main text. One hence concludes that the documented ability of the three fields to predict the location of plastic activity is not due to the choice of the descriptor. 

\section{Stability of the similarity analysis against grid size and interpolation scheme}\label{appstable}

\begin{figure*}[ht]
    \centering
    \includegraphics[width=1\linewidth]{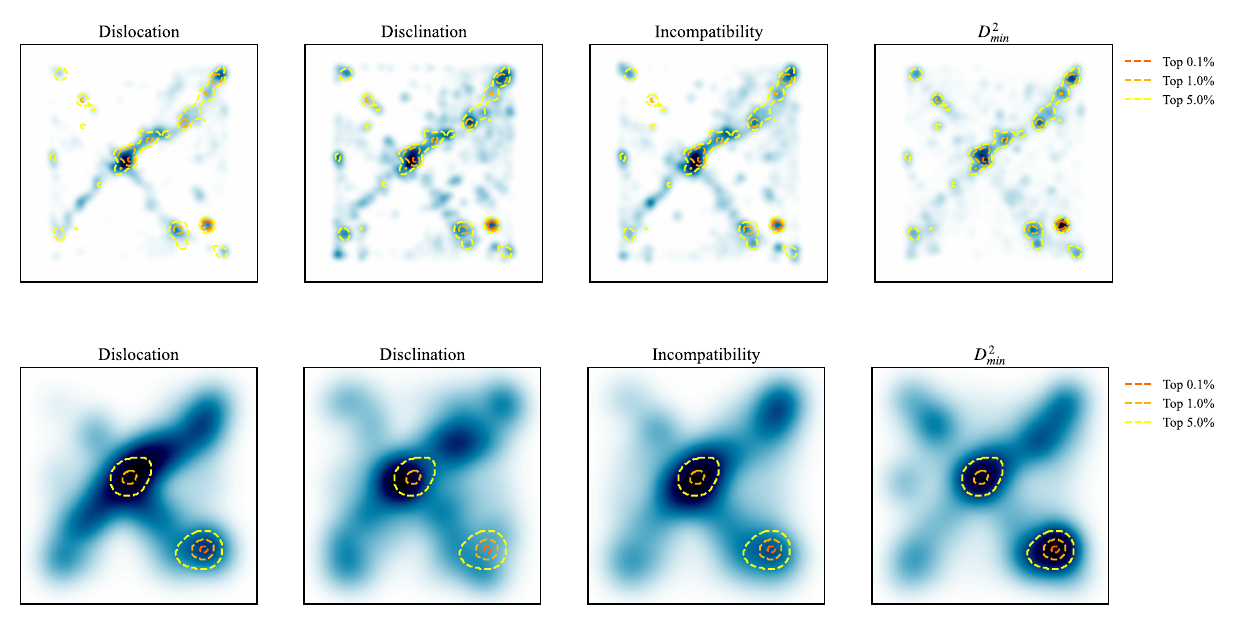}
    \caption{Influence of the grid size $a$ used for calculating the derivatives on the spatial distribution of the various fields. The system is the 2D experiment. \textbf{Top row: } Grid size $a = 0.5 \,d_s$. \textbf{Bottom row: }Grid size $a = 2.25 \,d_s$.}
    \label{fig:S4}
\end{figure*}

\begin{figure*}[ht]
    \centering
    \includegraphics[width=0.8\linewidth]{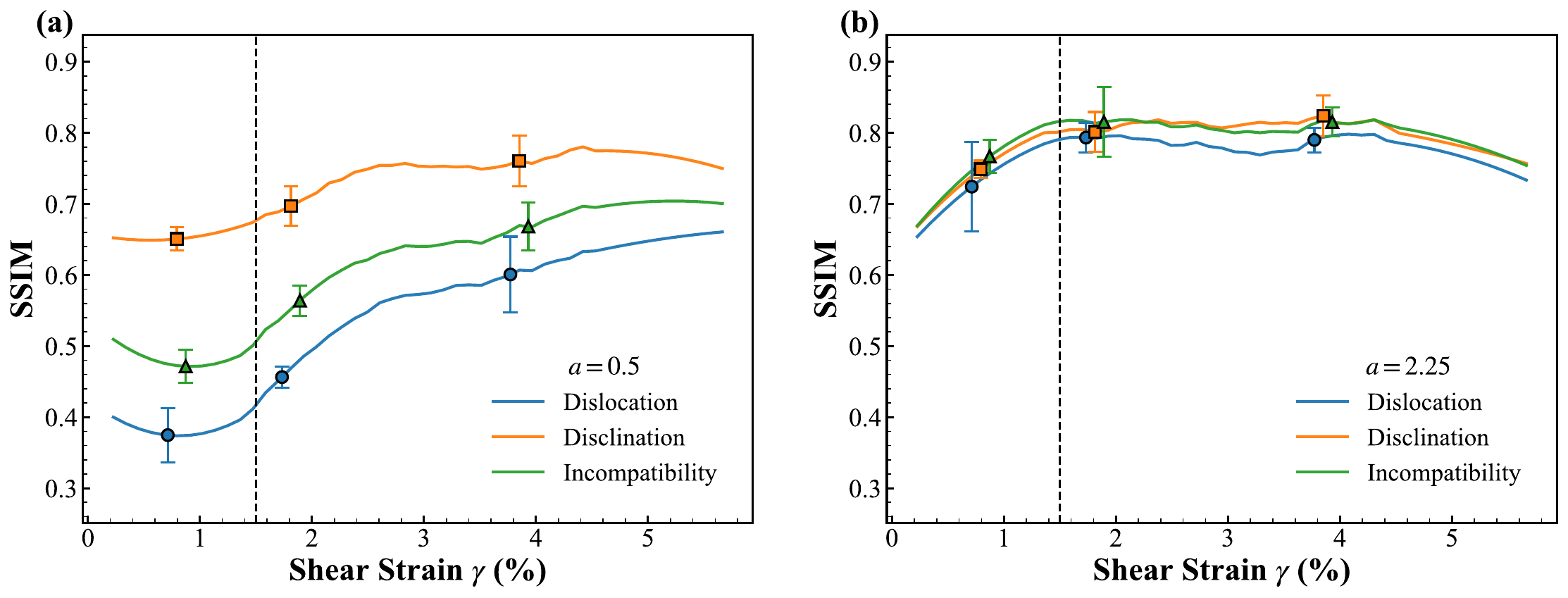}
    \caption{\textbf{(a)} Linear interpolation for 2D experiment: grid size $a = 0.5\, d_s$. \textbf{(b)} Linear interpolation and coarse grain for 2D experiment: grid size $a = 2.25 \,d_s$. 
    }
    \label{last}
\end{figure*}

In Figs.~\ref{fig:S4}, we demonstrate the robustness of our similarity analysis with respect to both the grid size and the interpolation scheme. The panels correspond to the 2D experimental data, with the specific parameters and methods indicated in each caption. The qualitative agreement across different choices is clear. Consistently, the SSIM analysis reported in Fig.~\ref{last} exhibits the same behavior, and becomes larger when the grid chosen is not too small.

\section{Defect densities for different strain intervals}\label{appstrain}
\begin{figure*}[h]
    \centering
    \includegraphics[width=\linewidth]{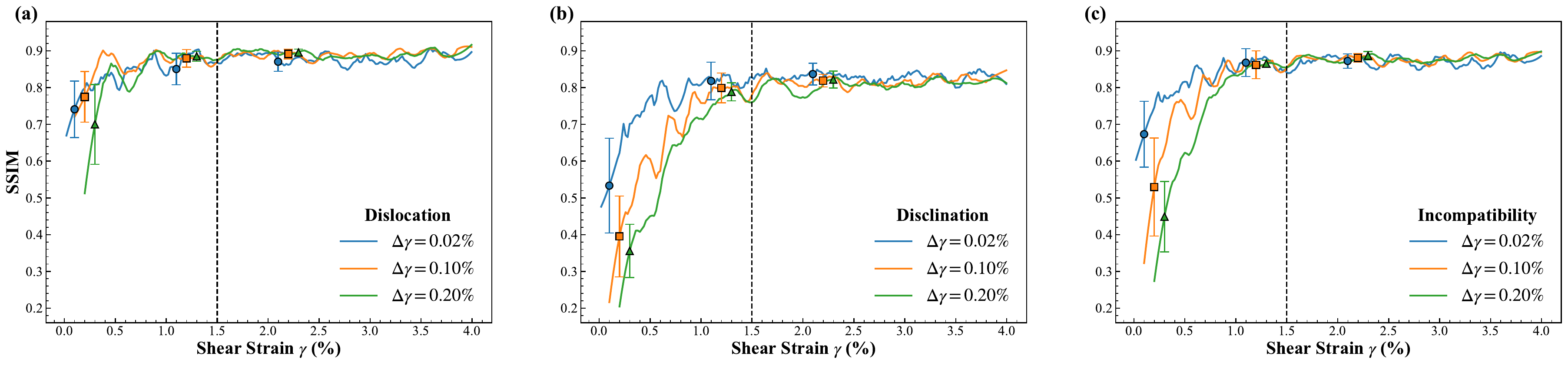}
    \caption{Structural Similarity Index Measure (SSIM) for different strain intervals, $\Delta \gamma = 0.02\%, 0.10\%$, and $0.20\%$, between $D^2_{\text{min}}$ and \textbf{(a)} the dislocation, \textbf{(b)} disclination, and \textbf{(c)} incompatibility density fields. These data refer to the 2D simulations. The vertical dashed lines indicate the yielding strain for each system.}
    \label{fig:diff_delta}
\end{figure*}

The details of the dislocation, disclination, and incompatibility fields will depend on the strain increment used to calculate the displacement field.
To assess the robustness of the analysis presented in the main text with respect to the choice of this strain interval $\Delta \gamma$, we examine two additional values, $\Delta \gamma = 0.1\%$ and $0.2\%$, using the 2D Lennard-Jones simulation data. The resulting values for the SSIM are shown in Fig.~\ref{fig:diff_delta}, where we include also the data from the main text, i.e., $\Delta \gamma=0.02$\%.).

Overall, the correlation between the defect density fields and $D^2_{\text{min}}$ remains robust, particularly above yielding (indicated by the vertical dashed black line). Interestingly, increasing the strain interval leads to a gradual deterioration of the correlation between $D^2_{\text{min}}$ and both the disclination and incompatibility densities for $\gamma \lesssim 1\%$. In contrast, the correlation between the dislocation density and $D^2_{\text{min}}$ appears largely insensitive to variations in $\Delta \gamma$.

\normalem
%

\end{document}